\documentclass[11pt]{article}
\usepackage{amsfonts,fullpage,graphicx,vmargin,natbib,mathrsfs}
\usepackage{pstricks}
\usepackage{subcaption}
\usepackage{amsmath}
\usepackage{amsthm}
\usepackage{multirow}
\usepackage{comment}
\usepackage{url}
\usepackage[inline]{enumitem}
\usepackage{placeins}

\newcommand{\iid}{\stackrel{\mbox{\scriptsize iid}}{\sim}}

\newcommand{\bm}[1]{\mbox{\boldmath{$#1$}}}

\newcommand{\calN}{\mathcal{N}}

\newcommand{\e}{\mathrm{e}}

\newcommand{\R}{\mathbb{R}}

\usepackage{mathtools}

\newcommand{\virgolette}[1]{`#1'}

\renewcommand{\mid}{\ensuremath{\,|\,}}

\usepackage[normalem]{ulem}
\newcommand{\change}[1]{{\color{black}{#1}}} %

\setmarginsrb{1in}{1.18in}{1in}{1.18in}{0pt}{0mm}{0pt}{0.4in}

\makeatletter
\begin{document}

\title{\bf MCMC computations for Bayesian mixture models using repulsive  point processes}
\author{  
Mario Beraha\thanks{Department of Mathematics, Politecnico di Milano, Milano, Italy}
  \thanks{Department of Computer Science, Universit\`a di Bologna, Bologna, Italy}, 
  Raffaele Argiento\thanks{Department of Statistical Sciences, Universit\`a Cattolica del Sacro Cuore, Milano, Milano, Italy},
  Jesper M{\o}ller\thanks{
  Department of Mathematical Sciences, Aalborg University, Aalborg, Denmark \newline
  Jesper M{\o}ller was supported by the `Danish Council for Independent Research -- Natural Sciences' grant DFF -- 7014-00074 `Statistics for point processes in space and beyond'
  },
  Alessandra Guglielmi\footnotemark[1]
}
\date{18 April 2021}
  
\maketitle

\bigskip
\begin{abstract}
Repulsive mixture models have recently gained popularity for Bayesian cluster detection. 
Compared to more traditional mixture models, repulsive mixture models produce
a smaller number of well separated clusters. The most commonly used methods for posterior
inference either require to fix a priori the number of components or are based on reversible
jump MCMC computation. We present a general framework for mixture models, when the 
prior of the `cluster centres' is a finite repulsive point process depending on a hyperparameter,
specified by a density which may depend on an intractable normalizing constant.
By investigating the posterior characterization of this class of mixture models, we derive a 
MCMC algorithm which avoids the well-known difficulties associated to reversible jump 
MCMC computation. In particular,  we use an ancillary variable method, which eliminates 
the problem of having intractable normalizing constants in the Hastings ratio.
The ancillary variable method relies on a perfect simulation algorithm, and we demonstrate
this is fast because the number of components is typically small. In several simulation studies
and an application on sociological data, we illustrate the advantage of our new methodology
over existing methods, and we compare the use of a determinantal or a repulsive Gibbs point process prior model.
	\end{abstract}
	
\noindent%

\section{Introduction}\label{sec:intro}

Mixture models are useful when partitioning observations $\bm y=(y_1,\dots,y_n)$ into groups/clusters as well as when 
approximating densities that are not otherwise modelled by standard parametric distributions; see \cite{fruhwirth2019handbook} and the references therein. 
Our approach to finite mixture models extends that in \cite{argiento2019infinity},  but the
focus in this paper will be on prior specification and Bayesian MCMC computations when the aim is cluster detection. 
Our objective is partly to present a general framework for mixture models based on repulsive point process priors for `cluster centres', arguing why this is useful, and partly to derive a MCMC algorithm which avoids the well-known
difficulties associated to reversible jump MCMC computation. 
In several
simulation studies and an application on sociological data, we illustrate the advantage of our new methodology over existing methods, and we compare the use of the different repulsive point process priors. 
Moreover, when introducing a hyperparameter in such priors, we demonstrate that perfect simulation is fast in connection to a useful ancillary variable method. 

\subsection{Setting}\label{sec:1.1}

For specificity, assume each $y_i \in  \R^q$ with $q \geq 1$. It will always be obvious from the context whether we consider $y_i$ (and other variables considered later on) as a random variable, a realization, or an argument of a function. Denote
$\{k(\cdot \,|\, \tau):\, \tau \in \Theta \}$ a parametric family of 
densities  
(with respect to Lebesgue measure on $\mathbb R^q$ or counting measure on a countable subset of $\mathbb R^q$), where the specification of the parameter space $\Theta$ is application dependent.
We refer to this parametric family as the kernel (of the mixture model). 
We consider each $y_i$ to follow a mixture of these densities: 
 Let $\bm\tau=(\tau_1,\dots,\tau_m)$ specify  $m$ densities 
where $\tau_h \in \Theta$ for $h=1, \ldots, m$, and let
$\bm {w}=(w_1,\dots,w_m)$  specify weights $w_h>0$ with $\sum_{h=1}^{m}w_h=1$. We assume that $\bm\tau, \bm w$, and $m\leq +\infty$  are random. The case where $m$ is a fixed positive integer may be considered as a special case. 
Note that $m$ is determined by $\bm\tau$ as well as by $\bm w$.
Conditioned on $(\bm{w},\bm\tau)$, the observations are assumed to be independent identically distributed (iid) with a distribution given by the following mixture density:
\begin{equation}
    y_i \iid \sum_{h=1}^m w_h k(\cdot \,|\, \tau_h),\qquad i=1,\ldots,n. 
    \label{eq:mixture}
\end{equation} 
The densities $k(\cdot\,|\,\tau_h)$, $h=1, \ldots, m$ are usually referred to
as the `components' of the mixture.  
In this context, cluster detection means estimating allocation parameters $\bm c=(c_1,\ldots,c_n)\in\{1,\ldots,m\}^n$ where the sets $\{y_i:\,c_i=h\}$, $h=1,\ldots,m$ are the clusters.  
The number of clusters in the mixture model is the number of allocated components in \eqref{eq:mixture}, i.e., the number of unique values in $(c_1,\ldots,c_n)$.

We make prior assumptions as follows. To control the number of clusters, $m$ is  
random and finite; the case $m= + \infty$ would be relevant for nonparametric   inference \citep{muller2013bayesian},  but this context is not addressed in this paper. 
Only when $m<+\infty$ is not fixed, it can be consistently estimated, cf.\ \cite{argiento2019infinity} and \cite{miller2018mixture}. 
 We let $\tau_h=(\mu_h,\gamma_h)$, thinking of $\mu_h$ as a continuous random  parameter in $\mathbb R^q$ which specifies a `cluster centre' of cluster $h$, and of $\gamma_h$ as a positive random parameter ($q=1$) or a continuous covariance matrix ($q\ge2$) (or, in simple settings, a fixed positive number) which specifies
the amount of dispersion of the data points in cluster $h$ (for example, $k(\cdot \,|\, \tau_h)$ could be a normal density with mean $\mu_h$ and variance $\gamma_h$).  
 To make posterior inference more robust, we add a hyperparameter $\xi$ to the prior distribution of $(\mu_1,\ldots,\mu_m)$. 
 Furthermore, since the mixture density in \eqref{eq:mixture} does not depend on the order of the components, we can assume that
 \begin{enumerate}
 \item[(a)] the conditional marginal prior density $p(\mu_1,\ldots,\mu_m\,|\,\xi, m)$ is exchangeable,
 \end{enumerate} 
  that is, for any fixed integer $m\ge1$, it is invariant under permutations of $\mu_1,\ldots,\mu_m$. Note that $\bm\mu=\{\mu_1,\ldots,\mu_m\}$ is then a finite point process, specifying both the random number $m$ of components and the locations of the cluster centres.    
Finally, a priori we make conditional independence assumptions:  Conditioned on $(\xi,m)$, we have that
  \begin{enumerate}
  \item[(b)]  $(w_1,\ldots,w_m)$,
 $(\mu_1,\ldots,\mu_m)$,  and $(\gamma_1,\ldots,\gamma_m)$ are a priori independent, 
 \item[(c)]  given $m$, the conditional marginal prior distribution of $(w_1,\ldots,w_m)$ does not depend on $\xi$,
 \item[(d)] the $\gamma_h$'s are iid, with a prior distribution which does not depend on $(\xi,m)$,
 \end{enumerate} 
 and conditioned on $(\xi,m,w_1,\ldots,w_m,\mu_1,\ldots,\mu_m,\gamma_1,\ldots,\gamma_m)$, we have that
 \begin{enumerate}
 \item[(e)]
 the $c_i$'s are iid with a prior distribution given by $P(c_i=h\,|\, \bm w)=w_h$.     
\end{enumerate}   
Hence,  the random parameter here consists of $( \xi,\{\mu_1,\ldots,\mu_m\},w_1,\ldots,w_m,\gamma_1,\ldots,\gamma_m,c_1,\ldots,c_n)$. By Bayes' theorem, using the generic notation $p(\cdot)$ for a density  and $p(\cdot\,|\,\cdot)$ for a conditional density, the posterior density becomes
 \begin{multline}\label{e:p}
 p(\xi,\{\mu_1,\ldots,\mu_m\},w_1,\ldots,w_m,\gamma_1,\ldots,\gamma_m,c_1,\ldots,c_n\,|\, y_1,\ldots,y_n)\propto \\
  p(\xi)p(m\mid\xi)p(\mu_1,\ldots,\mu_m\mid \xi,m)       %
  p(w_1,\ldots,w_m\,|\, m)\left[\prod_{h=1}^m p(\gamma_h)\right]\left[\prod_{i=1}^nw_{c_i}k(y_i\,|\, (\mu_{c_i},\gamma_{c_i}))\right].
 \end{multline} 
The dominating measure for \eqref{e:p} is given in Section~\ref{sec:posterior} which contains measure theoretical details;  see Section~\ref{sec:prior_mu} for further prior specifications. In brief, 
the prior specification of $\bm\mu$ and $\bm w$ requires particular attention, whilst for the prior specification of the remaining parameters we use a standard setting, following \cite{fraley2007bayesian}.    
 
\subsection{Previous work on repulsive mixture models}
\label{sec:previouswork}
 
The most used mixture models assume the $\mu_h$'s are iid and independent of $m$, cf.\ \cite{fruhwirth2019handbook}. This assumption,
 although convenient for mathematical tractability, is often an oversimplification and might produce misleading results in producing too many clusters. This motivated \cite{petralia2012repulsive}, \cite{xu2016bayesian}, \cite{fuquene2019choosing},  \cite{quinlan2017parsimonious}, \cite{bianchini2018determinantal}, and \cite{xie2019bayesian}  to explicitly define prior models with repulsion between the locations, thereby obtaining well separated components.  
 
In \cite{petralia2012repulsive}, \cite{fuquene2019choosing}, and \cite{quinlan2017parsimonious}, $m$ is finite and fixed, but, as mentioned before, this cannot guarantee posterior consistency of the number of components. 
However, 
\cite{xu2016bayesian}, \cite{bianchini2018determinantal}, and \cite{xie2019bayesian}   assumed $m$ to be finite and random. In particular, \cite{xu2016bayesian} and \cite{bianchini2018determinantal}
 dealt with determinantal point process (DPP) priors for $\bm\mu$. A DPP density has to be approximated as described in \cite{lavancier2015determinantal} where the calculation will increase \change{exponentially} fast as the dimension $q$ increases,  or as described in \cite{NIPS2015_5810} but at the price that the model parameters are hard to interpret. 
 \cite{xie2019bayesian}  had no hyperparameter $\xi$ in their repulsive mixture model,  which used as a prior for $\bm\mu$ conditioned on $m$, a tempered repulsive pairwise interaction point process density 
of the form
  \begin{equation*}
  \label{e:crazy}
  p(\mu_1,\ldots,\mu_m\,|\,m)=\frac{1}{Z_m}\left[\prod_{i=1}^m\phi_1(\mu_i)\right]\left[\prod_{1\le i<j\le m}\phi_2(\|\mu_i-\mu_j\|)^{1/m}\right]\end{equation*}
  with respect to $m$-fold Lebesgue measure on $\mathbb R^q$. Here, $\|\cdot\|$ denotes usual distance, $\phi_1$ is a non-negative function, $0\le\phi_2\le1$ is a \change{non-decreasing} function (this implies repulsiveness), and  $Z_m$ is the normalizing constant.  Note that if $\phi_2(\cdot)=1$, then $\mu_1,\ldots,\mu_m$ are iid and independent of $m$. Apart from this case, $Z_m$
 is intractable and has to be approximated by numerical methods,
a non-trivial task which limits both efficiency and feasibility as $q$ increases.

As far as posterior simulation is concerned, \cite{xu2016bayesian} and \cite{bianchini2018determinantal} proposed to simulate $(\bm w,\bm\tau)$
using a reversible jump MCMC algorithm, cf.\ \cite{green1995reversible}.
At every iteration of this algorithm, either a split move (in which one component is
killed and two new ones are created, hence increasing the dimension by one), or a combine move
(in which two components are merged into a single one, hence decreasing the dimension by one)
is proposed.
As discussed in \cite{green2003trans}, \cite{richardson1997bayesian}, and \cite{dellaportas2006multivariate}, in order to obtain good mixing properties 
of the reversible jump MCMC algorithm, it is crucial
to define appropriate proposal distributions that generate the new values
in the split move. In general, this is a complex task that depends heavily 
on the kernel under consideration. 
  
Similarly to how \cite{miller2018mixture} studied a classical mixture model, \cite{xie2019bayesian} considered in the observation model \eqref{eq:mixture} to marginalize with respect to a prior of $(\bm w,\bm\tau)$ and derived a
`marginal MCMC algorithm'.
However, although this algorithm compared with the reversible jump MCMC algorithm has smaller 
auto-correlations for the number of clusters, it requires the calculation of the normalizing constants $Z_1,Z_2,\ldots$ up to some truncation, and inference is limited to  the number of clusters 
and the posterior mean of the mixture density.

\subsection{Our contribution and outline}
\label{sec:our_cont_out}

We discuss a general framework  for mixture models based on repulsive point process priors for `cluster centres' $\bm \mu$ and derive a new MCMC algorithm avoiding  the problem with reversible jump MCMC computation.

Our first contribution is the proposal of the prior of $\bm\mu$ conditioned on $\xi$, cf.\  item (a) in Section~\ref{sec:1.1}: We consider a general setting with a repulsive finite point process density, including the case of a DPP (any DPP except the special case of a Poisson process is repulsive) or a density specified by an unnormalized density,
e.g.\ a pairwise interaction point process density, which involves a normalizing constant $Z_\xi$ which in general (except the special case of a Poisson process) is intractable. 
As a particular simple example of a pairwise interaction point process, we  assume a Strauss process (defined later in Section~\ref{sec:rep-p-int}). 
Note that the prior distributions for $\bm\mu$ in all the papers cited in Section~\ref{sec:previouswork} can all be considered as special cases of our prior for $\bm\mu$. 
Notice also that $Z_\xi$ will never appear in our posterior simulation algorithm.  

The second contribution is the algorithm for posterior simulation from our model. This contribution builds upon \cite{argiento2019infinity} and is mainly based on two assumptions, namely $\bm w$ and $\bm \mu$ are chosen a priori independent and the mixture weights $\bm w$ are defined by \emph{normalization} of iid infinitely divisible random variables, i.e. $\bm w$ follows a  {\it normalized infinitely divisible} distribution \citep{favaro2011class}. 
Indeed, \cite{argiento2019infinity} introduced the class of \emph{normalized independent point processes mixture} models and showed that this class can be framed into the Bayesian nonparametric context. In this way, several ideas and algorithms developed in the nonparametric literature for normalized random measures with independent increments \citep[NRMI -- see][]{regazzini2003distributional} 
can be adapted to the finite dimensional case.
Here we extend \cite{argiento2019infinity}  building a Metropolis-within Gibbs sampler, referred to as \emph{conditional Gibbs sampler} in the Bayesian nonparametric  literature, see \cite{papaspiliopoulos2008retrospective}.
In particular, we relax the usual assumption of $\mu_1,\ldots,\mu_m$ being iid and independent of $m$, still being able to propose a transformation of $\bm\mu$ into allocated cluster centres $\bm\mu^{(a)}=\{\mu_{c_i}:\,i=1,\ldots,n\}$ and non-allocated cluster centres $\bm\mu^{(na)}=\bm\mu\setminus\bm\mu^{(na)}$. This allow us to simulate from the full conditional of $\bm\mu$ without resorting to
the split and combine moves of the reversible jump MCMC algorithm as used in \cite{xu2016bayesian} and \cite{bianchini2018determinantal}. In fact posterior updates of $\bm\mu^{(a)}$ becomes easy and when updating $\bm\mu^{(na)}$ we use the Metropolis-Hasting birth-death algorithm in \cite{geyer1994simulation}. 
The Metropolis-Hasting birth-death algorithm  
has the  advantage that the choice of the kernel does not impact on the acceptance rate of the algorithm.

We impose the hyperprior on $\xi$, the parameter in the repulsive point process prior controlling the intensity of the point process, to make posterior inference more robust, cf.\ Section~\ref{sec:1.1}, unlike previous literature \citep[apart from][]{bianchini2018determinantal}.
When making posterior updates of $\xi$ in our MCMC algorithm, if the prior density for $\bm\mu$ conditioned on $\xi$ has an intractable normalizing constant $Z_\xi$, we get rid of $Z_\xi$ mentioned by using the single exchange algorithm in \cite{murray2006mcmc} coming from the ancillary variable algorithm in \cite{moller2006efficient}. These algorithms require perfect simulation of an ancillary variable following the same distribution as $\bm\mu$ conditioned on $\xi$. Interestingly, perfect simulation is feasible in our context because $m$ will typically be small (in our examples it is effectively always less than 10). 

The remainder of this paper is organized as follows. Sections~\ref{sec:prior_mu} and \ref{sec:prior_weights} specify our further prior assumptions on the cluster centers $\bm \mu$ and the mixture weights $\bm w$, respectively. Section~\ref{sec:posterior} derive the posterior density, using the useful superposition of $\bm \mu$ mentioned  above, and provides the technical details needed when dealing with point process densities (we aim at keeping this as simple as possible). Section~\ref{sec:algorithm} details our Metropolis-within-Gibbs \change{sampler} for posterior simulation.
Sections~\ref{sec:strauss_params} and \ref{sec:dpp_params} discuss prior elicitation when the prior for $\bm \mu$ is the Strauss process  and the DPP (conditioned on $\{m\ge 1\}$, see Section~\ref{sec:prior_mu}).
Section~\ref{sec:simulation} presents various simulation studies comparing posterior inference and MCMC mixing obtained using reversible jump or our \change{Metropolis-within-Gibbs sampler}, and using a DPP, a Strauss process, or a non-repulsive prior for $\bm\mu$. Furthermore, an application to \change{a} sociological data set is discussed in Section~\ref{sec:teenage}. The article concludes with a discussion in Section~\ref{sec:discussion}.
\change{In the Appendix we provide practical details on our Metropolis-within-Gibbs sampler, collects additional simulation studies, including an illustration on the advantages of using a Strauss
process over a DPP as prior for $\bm \mu$, and discusses possible extensions.
The code for posterior simulation has been implemented in C++ and linked to Python.}

\section{Prior specification of $\bm\mu$}
\label{sec:prior_mu}

For the prior specification of $\bm\mu$, introduced in item (a) in Section~\ref{sec:1.1}, which is the first original contribution of our work, a few technical details are needed to start.  
Let $\Omega=\cup_{m=0}^\infty\Omega_m$ denote the space of all finite subsets (point configurations) of $\mathbb R^q$, where $\Omega_m$ denotes the space of all finite subsets of cardinality $m$, with $\Omega_0=\{\emptyset\}$,  
  where $\emptyset$ denotes the empty point configuration; although we cannot have 0 groups, it becomes convenient in Section~\ref{sec:posterior} to include $\Omega_0$ into the definition of $\Omega$. 
We equip each $\Omega_m$ with the smallest $\sigma$-algebra making the mapping of pairwise distinct $(\mu_1,\ldots,\mu_m)\in\mathbb R^{qm}$ into $\{\mu_1,\ldots,\mu_m\}\in\Omega_m$ measurable. The $\sigma$-algebra on $\Omega$ is the smallest $\sigma$-algebra containing the union of the $\sigma$-algebras on each $\Omega_m$. 
  Then $\bm\mu$
 is absolutely continuous with respect to a measure on $\Omega$  which, with an abuse of notation, is denoted $\mathrm d\bm\mu$ and defined as follows. For sets $B=\cup_{m=0}^\infty B_m$ with $B_m\subseteq\Omega_m$,
\[
 \int_B\mathrm d\bm\mu=\sum_{m=0}^\infty\frac{1}{m!}\int_{B_m}\,\mathrm d\bm\mu_m,
\]
where the notation means the following. For $m=0$,
 we interpret
 the term in the sum as $\mathbb[\emptyset\in A]$.
We set $\bm\mu_m=\{\mu_1,\ldots,\mu_m\}$ and, with an abuse of notation, write $\mathrm d\bm\mu_m$ for Lebesgue measure $\mathrm d\mu_1\cdots\mathrm d\mu_m$ on $\mathbb R^{qm}$.  
Further, we write $\int_{B_m}\,\mathrm d\bm\mu_m$ for $\int_{\mathbb{R}^{qm}} \mathbb{I}[\bm \mu \in B_m]\,\mathrm d\bm\mu_m$,
where $\mathbb I[\cdot]$ denotes the indicator function.
 Then, 
conditioned on $\xi$, the density of $\bm\mu$ with respect to $\mathrm d\bm\mu$ is given by
 \[p(\bm\mu\,|\,\xi)=  p(m\,|\,\xi)p(\mu_1,\ldots,\mu_m\,|\,\xi, m),\qquad \bm\mu=\{\mu_1,\ldots,\mu_m\}\in\Omega,\ m\ge1,\]
  setting  $p(\mu_1,\ldots,\mu_m\mid\xi,m)=0$ if $m=0$.
This means that we consider the prior process prior restricted to the event that $\bm\mu$ is non-empty.

\subsection{Repulsive pairwise-interaction point process priors}\label{sec:rep-p-int}

When incorporating repulsiveness in the prior density $p(\bm\mu\mid\xi)$, we suggest a repulsive pairwise-interaction point process. 
This is 
a popular class of models in statistical physics and spatial statistics; see \cite{moller2003statistical} and the references therein. 
The repulsive pairwise-interaction density is of the form
\begin{equation}\label{e:pairwise}
p(\bm\mu\mid\xi)=\frac{1}{Z_\xi}\left[\prod_{h=1}^m\phi_1(\mu_h\mid\xi)\right]\left[\prod_{1\le i<j\le m}\phi_2(\|\mu_i-\mu_j\|\mid\xi)\right],
\end{equation}
where $\phi_1(\cdot\mid\xi)\ge0$ is an integrable function,
$0\le\phi_2(\cdot\mid\xi)\le1$ is a \change{non-decreasing} function, and $Z_\xi$ is a normalizing constant. Note that $Z_\xi<+\infty$, but in general $Z_\xi$ is intractable. An exception is the special case $\phi_2(\cdot\mid\xi)=1$ (a Poisson process with intensity function 
$\phi_1(\cdot\mid\xi)$ and conditioned on not being empty), where $Z_\xi=1-\exp(-\int\phi_1(\mu_h\mid\xi)\,\mathrm d\mu_h)$. 

For simplicity and specificity, in Sections~\ref{sec:simulation}-\ref{sec:teenage}, we follow \cite{bianchini2018determinantal} in letting $\xi$ be a positive random variable and using an empirical Bayesian approach with 
\begin{equation}
\phi_1(\mu_h\mid\xi)=\xi\,\mathbb I[\mu_h \in R],
\label{eq:DPP_phi1}
\end{equation}
 where 
$R\subset\mathbb R^q$ is the smallest rectangular region containing the data $\bm y$ and with sides parallel to the usual axes in $\mathbb R^q$ (they advocate the use of this choice over other more complicated situations). 

The simplest non-trivial case is a Strauss prior,
\begin{equation}
    \phi_2(r\mid\xi)=\alpha^{\mathbb I[r\le \delta]},
\label{eq:DPP_phi2}
\end{equation}
so that $\xi$ enters only in the expression of $\phi_1$. 
Here, $\delta>0$ is a fixed parameter, specifying the range of interaction, and $0\le\alpha\le1$ is a fixed interaction parameter.  
Note that, if $\alpha=0$, we set $0^0=1$ and obtain a so-called hard core point process. If $\alpha=1$, we obtain a model with no interaction which is like a Poisson process except that we condition on that $\bm\mu$ is non-empty. 

\subsection{Repulsive priors specified by an unnormalized density} 

In the following we consider a 
general prior model given by 
\begin{equation}\label{e:general}
p(\bm\mu\mid\xi)=\frac{1}{Z_\xi}g(\bm\mu\mid\xi),
\end{equation}
where \change{$g(\cdot\mid\xi)$ is a so-called unnormalized density, meaning that $g(\cdot\mid\xi)$ is a non-negative measurable function such that the normalizing constant $Z_\xi$ is finite. Note that by assumption  $g(\emptyset\mu\mid\xi) = 0$.}
Specific examples of \eqref{e:general} can be found in \cite{moller2003statistical} and the references therein. 
In our simulation study and application example (Sections~\ref{sec:simulation}-\ref{sec:teenage})
we focus on the Strauss prior and a specific DPP prior given below, but considering \eqref{e:general} is useful in order to give a general exposition of our methodology. 

To describe interaction in the general model \eqref{e:general}, one possibility is to assume that for any $\bm\mu\in\Omega$ and $\mu^*\in\mathbb R^q\setminus\bm\mu$ we have $g(\bm\mu\cup\{\mu^*\}\mid \xi)>0\Rightarrow g(\bm\mu\mid\xi)>0$, and then consider the so-called Papangelou conditional intensity defined by
\[\lambda(\mu^*,\bm\mu\mid\xi):=g(\bm\mu\cup\{\mu^*\}\mid\xi)/g(\bm\mu\mid\xi)\]
(taking $0/0:=0$). Then we have repulsiveness if $\lambda(\mu^*,\bm\mu\mid\psi)$ is a non-increasing function of $\bm\mu$, that is, $\lambda(\mu^*,\bm\mu\mid\psi)\ge\lambda(\mu^*,\bm\mu\cup\{\mu'\}\mid\psi)$ for any $\mu'\in\mathbb R^q\setminus\bm\mu\cup\{\mu^*\}$, where inequality can not be replaced by an identity. Clearly, this is true for \eqref{e:pairwise} when $\phi_2(\cdot\mid\xi)\not=1$. 

\subsection{Determinantal point process priors}

A DPP density (conditioned on that the DPP is non-empty) is a special case of \eqref{e:general} but with repulsion characterized in another way than above \citep{hough2009zeros,lavancier2015determinantal,biscio2016quantifying,moller2018couplings}.
To work with a DPP density, we consider a compact region $R\subset\mathbb R^q$ with $\int_R\mathrm dx>0$, and a complex covariance function $C:R\times R\mapsto\mathbb C$ with a spectral representation
\begin{equation}\label{e:C-def}
C(x,x'\mid\xi)=\sum_{i=1}^\infty \lambda_i\varphi_i(x)\overline{\varphi_i(x')},\qquad x,x'\in R,
\end{equation}
where the $\varphi_i$'s form an orthonormal basis for the $L^2(R)$-space of complex functions defined on $R$, each $\lambda_i\ge 0$, and 
$\sum_{i=1}^\infty \lambda_i<+\infty$. 
Then existence of the DPP is equivalent to that all $\lambda_i\le1$, cf.\ \cite{Macchi:75}. Note that we suppress in the notation that the eigenvalues $\lambda_i$'s and the eigenfunctions $\varphi_i$'s may depend on $\xi$. 

A special case of a DPP occurs when $C$ is a projection of finite rank $m$, let us say 
\[C(x,x'\mid\xi)=\sum_{i=1}^m \varphi_i(x)\overline{\varphi_i(x')},\qquad x,x'\in R.\]
From \eqref{e:C-def} we obtain a density 
\begin{equation}\label{e:DPPprojection}
p(\mu_1,\ldots,\mu_m\mid\xi)={\mathrm{det}}\{C(\mu_h,\mu_h')\}_{h,h'=1,\ldots,m} \qquad\mbox{for  $\mu_1,\ldots,\mu_m\in R$}, 
\end{equation}
where ${\mathrm{det}}\{C(\mu_h,\mu_h')\}_{h,h'=1,\ldots,m}$ is the determinant of the $m\times m$ matrix $\{C(\mu_h,\mu_h')\}_{h,h'=1,\ldots,m}$. 
A point process with density \eqref{e:DPPprojection} 
is called a projection DPP with kernel $C$. Note that it consists of exactly $m$ points in $R$.

The general construction of a DPP is given by introducing a random projection 
\begin{equation}\label{e:K}
K(x,x'\mid\xi)=\sum_{i=1}^\infty B_i\varphi_i(x)\overline{\varphi_i(x')},
\end{equation} 
where $B_1,B_2,\ldots$ are independent Bernoulli variables with means $\lambda_1,\lambda_2,\ldots$, respectively. Then a DPP with kernel $C$ is a finite point process on $R$ which conditioned on $B_1,B_2,\ldots$ is a projection DPP with kernel $K$; it can be shown that the distribution of this DPP depends only on $C$, cf.\ \cite{Hough:etal:06}. Note that $\sum_{i=1}^\infty B_i$ is the random number of points. In particular, assuming all $\lambda_i<1$ and defining $C'$ as $C$ in \eqref{e:C-def} but with each $\lambda_i$ replaced by $\lambda_i'=\lambda_i/(1-\lambda_i)$, the DPP 
  has unnormalized density
\begin{equation}\label{e:unnormdppdensity}
g(\bm\mu\mid\xi)={\mathrm{det}}\{C'(\mu_h,\mu_{h'})\}_{h,h'=1,\ldots,m} \qquad\mbox{for pairwise distinct $\bm\mu=\{\mu_1,\ldots,\mu_m\}\subset R$}, m\ge1. 
\end{equation}

Most DPP densities are specified as in \eqref{e:unnormdppdensity} with the kernel coming from 
a  parametric family of (often real) covariance functions with all eigenvalues $<1$, see \cite{lavancier2015determinantal}.
The advantage of using such models is that we can avoid \change{including} the Bernoulli variables as ancillary variables in the posterior, whilst 
the problem is to find a spectral representation. Note that when we condition on that the DPP is non-empty,
the normalizing constant is given by
\begin{equation}\label{e:norm-const}
Z_\xi=\change{\prod_{i=1}^\infty(1-\lambda_i)^{-1} - 1}.
\end{equation}

For our purpose it is easiest to let $R$ be rectangular and use a spectral approach with Fourier basis functions for the eigenfunctions, cf.\ \cite{lavancier2015determinantal}. In Sections~\ref{sec:simulation}-\ref{sec:teenage}, we follow \cite{bianchini2018determinantal} in making this choice of eigenfunctions and letting $R=[-\tfrac12,\tfrac12]^q$ and
\begin{equation}\label{e:C}
C(x,x'\mid\xi)=\sum_{j\in\mathbb Z^q}\lambda_j\cos(2\pi j\cdot(x-x')),
\end{equation}
where $\mathbb Z$ is the set of integers, $\cdot$ denotes the usual inner product on $\mathbb R^q$, and $\lambda_j=\chi(j)$
is specified by the spectral density $\chi$ of the power exponential spectral model  from \cite{lavancier2015determinantal}. Specifically, 
\begin{equation}\label{eq:spectral}
    \lambda_j=\xi \frac{\alpha^q \Gamma(q/2 + 1)}{\pi^{q/2} \Gamma(q / \beta + 1)} \exp(- \|\alpha j\|^\nu),
\end{equation}
where $\Gamma(\cdot)$ is the gamma function, $\alpha$ and $\beta$ are fixed positive parameters, and $\lambda_j$ depends on the parameter 
$\xi>0$ so that $\lambda_j \le 1$  and $\sum_{j \in \mathbb{Z}^q} \lambda_j<\infty$. For details, see \cite{lavancier2015determinantal}, noting that $\xi$ is the intensity 
of the DPP if we do not condition on that the DPP is non-empty.

When dealing with computations, in the sum of \eqref{e:C} and in the corresponding product $\prod_{j\in\mathbb Z^q}\cdots$ for the normalizing constant, cf.\ \eqref{e:norm-const}, we may replace the infinite lattice $\mathbb Z^q$ with a finite set, which is most naturally given by $\{-N,-N+1,\ldots,0,\ldots,N-1,N\}^q$, where $N>0$ is an integer. Then $m\le (2N+1)^q$; in \cite{bianchini2018determinantal}, $N=50$ for $q=1,2$. \cite{NIPS2015_5810} suggested
an alternative approach, which does not require the spectral approach used above but specifies the DPP density directly by \eqref{e:unnormdppdensity} and exploits certain bounds for the product in \eqref{e:norm-const}. However, it is then harder to interpret the parameters, and in particular to work with an intensity parameter.

 In order to fix the values of $\alpha$ and $\beta$, we could follow \cite{lavancier2015determinantal} who proposed to approximate some summaries such as the pair correlation function which depends only on $(\alpha,\beta)$.  Instead, in Section~\ref{sec:prior}, we discuss an empirical Bayesian approach to select hyperparameters and hyperpriors for both the Strauss process given by \eqref{e:pairwise}-\eqref{eq:DPP_phi2} and the DPP given by \eqref{e:C}-\eqref{eq:spectral}.

\section{Normalized infinite divisible prior for the weights}
\label{sec:prior_weights}

When deriving full conditional distributions for our Metropolis-within-Gibbs sampler given in Section~\ref{sec:algorithm}, it becomes convenient to introduce ancillary variables $t$ 
and $u$ as specified below. 

Conditioned on $m$, let $\bm s=(s_1,\ldots,s_m)$ consists of iid positive continuous random variables, with the distribution of each $s_h$ not depending on $m$, and with $\bm s$ independent of $(\xi,\bm\mu,\bm\gamma)$. Set $t=\sum_{h=1}^ms_h$ and $\bm w=(s_1/t,\ldots,s_m/t)$, so $\bm s$ and $(\bm w,t)$ are in a one-to-one correspondence.  In particular, in Sections \ref{sec:simulation}-\ref{sec:teenage}, we assume  each $s_h$ follows a gamma distribution, in which case our model 
can be referred to as a \textit{finite Dirichlet mixture model with repulsive locations}. We point out that the idea of building the weights $\bm w$ by normalization not only has computational advantages -- as discussed in Section~\ref{sec:our_cont_out} -- but it also allows us to embed the model into the large class of mixtures obtained by normalization of finite point processes \citep{argiento2019infinity}. This latter class, to be defined, requires only the distribution of $s_h$'s to be infinitely divisible, and it is the finite-dimensional counterpart of the normalized random measures with independent increments, 
which has been thoroughly investigated in the last two decades in the Bayesian nonparametric literature \citep[see, for instance,][]{regazzini2003distributional, james2009posterior, lijoi2010models}. 
 The weights $\bm w$ resulting from a finite normalization  have distribution on the simplex that is denominated as \textit{normalized infinite divisible} following \cite{favaro2011class}. 
 It is worth underlining that our \change{Metropolis-within-Gibbs sampler}, cf.\ Section~\ref{sec:algorithm}, works also for normalized infinite divisible priors different from the Dirichlet distribution, such as those introduced in \cite{argiento2019infinity}.

	One of the advantages of building the distribution of the weights $\bm{w}$ by normalization is that computations are easier. The main idea is to consider a gamma random variable $v$
	with scale parameter one and shape parameter $n$, which is independent of $(\xi,\bm \mu,\bm\gamma,\bm s,c)$. Then, we set the ancillary variable $u := v/t$. It is immediate to show that $u$ is well defined, i.e., it  has a density with respect to Lebesgue measure given by
\begin{equation*}
	p(u)=\frac{u^{n-1}}{\Gamma(n)}\int_0^\infty t^n \e^{-ut} p(t)dt
\end{equation*}
where $p(\cdot)$ in the integral is the density function of $t$. We show below (see \eqref{eq:u}) that, conditioned on $u$, the full conditional of the unnormalized weights $s_h$'s factorize (i.e., the weights are conditionally independent), so that simulation will be drastically simplified. We notice that the \emph{trick} of introducing the ancillary variable $u$ is familiar in the context of normalized completely random measure as mixing measures for mixture models.  It was studied in \cite{james2009posterior} in the infinite dimensional case and largely exploited by \cite{argiento2019infinity} and \cite{argiento2016posterior} in the finite dimensional setting.

\section{Posterior distribution and a useful decomposition of $\bm \mu$}

\label{sec:posterior} 

To specify the posterior obtained by considering all parameters introduced so far, including $(\bm s, t, u)$, we first notice that the dominating measure implicitly used in \eqref{e:p} leads to a new dominating measure $\nu$ given as follows. Let $\Xi$ and $\Gamma$ denote the spaces where $\xi$ and each $\gamma_h$ take values, respectively, equipped with some appropriate $\sigma$-algebras and measures $\mathrm d\xi$ and $\mathrm d\gamma_h$ (typically Borel $\sigma$-algebras and Lebesgue measures). 
For $m=1,2,\ldots$, set $\bm s_m=(s_1,\ldots,s_m)$ and $\bm\gamma_m=(\gamma_1,\ldots,\gamma_m)$, let 
$\mathrm d\bm s_m$ denote Lebesgue measure on $\mathbb R_+^m$, let $\mathrm d\bm\gamma_m$ denote the product measure $\prod_{h=1}^m\mathrm d\gamma_h$, and consider arbitrary measurable subsets $A\subseteq\Xi$, $B_m\subseteq \Omega_m$, $C_m\subseteq \mathbb R_+^m$, $D_m\subseteq \Gamma^m$, $E_m\subseteq\{1,\ldots,m\}^n$, and $F\subseteq\mathbb R_+$ (we still let the $\sigma$-algebra of $\Omega_m$ 
be induced by the Borel $\sigma$-algebra of $\mathbb R^{qm}$ and the mapping  $\mathbb R^{qm}\ni(\mu_1,\ldots,\mu_m)\mapsto\{\mu_1,\ldots,\mu_m\}\in\Omega_m$ with $\mu_1,\ldots,\mu_m$ pairwise distinct). 
Then the measure $\mathrm d\bm\mu$ together with the other reference measures lead to 
\begin{multline}\label{e:nu}
\nu(A\times \left\{\cup_{m=0}^\infty B_m\times C_m\times D_m\times E_m\right\}\times F)\\
=\int_A \,\mathrm d\xi\,
\sum_{m=0}^\infty\frac{1}{m!}
\int_{B_m}\,\mathrm d\bm\mu_m\, 
\int_{C_m} \,\mathrm d\bm s_m\,\int_{D_m} \,\mathrm d\bm\gamma_m\,
\sum_{c_1,\ldots,c_n=1}^m\mathbb I[\bm c\in E_m]
\, \int_{F}\,\mathrm du.
\end{multline} 
The posterior density  of the new parameter $(\xi,\bm\mu,\bm s,\bm\gamma,\bm c, u)$ with respect to $\nu$ is then
\begin{equation}\label{eq:post}
 p(\xi,\bm\mu,\bm s,\bm\gamma,\bm c, u \mid \bm y)\propto 
  p(\xi)p(\bm\mu\,|\, \xi)
  \left[\prod_{h=1}^m p(\gamma_h) p(s_h)\right] p(u\mid t) \frac{1}{t^n}  \left[\prod_{i=1}^n s_{c_i} k(y_i\,|\, (\mu_{c_i},\gamma_{c_i}))\right].
\end{equation} 

In the algorithm for posterior simulation presented in Section~\ref{sec:algorithm}, we find it useful to split $\bm\mu$ into those cluster centres which are
 used to allocate the data, and those which are not, that is,
$\bm\mu^{(a)} = \{\mu_{c_1}, \ldots, \mu_{c_n}\}$ and $\bm \mu^{(na)} = \bm \mu \setminus \bm \mu^{(a)}$. For the points of these processes we use the notation 
$\bm\mu^{(a)} =
\{\mu^{(a)}_1, \ldots, \mu^{(a)}_k\}= $ and $\bm \mu^{(na)} = \{\mu^{(na)}_1, \ldots, \mu^{(na)}_\ell\}$. Note that $1\le k \le m$, $\ell\ge0$, and the product measure $\mathrm d\bm\mu\times \mathrm d\bm\mu$ on $\Omega\times\Omega$ lifted by the map $(\bm x,\bm z)\mapsto \bm x\cup\bm z$ results in the measure $\mathrm d\bm\mu$. Hence, $(\bm\mu^{(a)},\bm\mu^{(na)})$ conditioned on $\xi$ has density
\[
    p(\bm\mu^{(a)},\bm\mu^{(na)}\mid\xi)=p(\bm\mu^{(a)}\cup\bm\mu^{(na)}\mid\xi)
\]
with respect to the product measure $\mathrm d\bm\mu^{(a)}\times \mathrm d\bm\mu^{(na)}$ (thinking of the measures $\mathrm d\bm\mu,\mathrm d\bm\mu^{(a)},\mathrm d\bm\mu^{(na)}$ as being identical but of course not thinking of $\bm\mu,\bm\mu^{(a)},\bm\mu^{(na)}$ as being identical).

Obviously, $(\bm\mu,\bm s,\bm\gamma,\bm c)$ and 
$(\bm\mu^{(a)},\bm s^{(a)},\bm\gamma^{(a)},\bm\mu^{(na)},\bm s^{(na)},\bm\gamma^{(na)},\bm c)$
are in a one-to-one correspondence, and the cardinalities of the point processes $\bm\mu^{(a)}$ and $\bm\mu^{(na)}$ satisfy $1\le k<+\infty$ and $0\le\ell<+\infty$. Finally,
setting $n_h=\#\{i:\,c_i=h\}$, we obtain from \eqref{e:nu} and \eqref{eq:post} the posterior density
\begin{equation}\label{eq:post-new}
\begin{aligned}
 p(\xi,\bm\mu^{(a)},\bm s^{(a)},&\bm\gamma^{(a)},\bm c,\bm\mu^{(na)},\bm s^{(na)},\bm\gamma^{(na)}, u \mid \bm y)\propto \\
 & p(\xi)p(\bm\mu^{(a)}\cup\bm\mu^{(na)}\,|\, \xi)
\left[\prod_{h=1}^k p(\gamma^{(a)}_h) p(s^{(a)}_h)(s^{(a)}_h)^{n_h}\prod_{i: c_i = h} k(y_i\,|\, (\mu^{(a)}_h,\gamma^{(a)}_h))\right]\\
&\hspace{6.5cm} \times \left[\prod_{h=1}^\ell p(\gamma^{(na)}_h) p(s^{(na)}_h)\right] p(u \mid t) \frac{1}{t^n}  
\end{aligned}
\end{equation} 
with respect to a new dominating measure defined by (using an obvious notation)
\begin{equation}\label{e:nu-new}
\begin{aligned}
&\nu'(A\times \left\{\cup_{k=1}^\infty B_k^{(a)}\times  C_k^{(a)}\times D_k^{(a)}\times E_k^{(a)}\right\}\times \left\{\cup_{\ell=0}^\infty B_\ell^{(na)}\times C_\ell^{(na)}\times D_\ell^{(na)}\right\}\times F)\\
&\hspace{2cm}= \int_A \,\mathrm d\xi\,
\sum_{k=1}^\infty\frac{1}{k!}
\int_{B_k^{(a)}}\,\mathrm d\bm\mu_k^{(a)}\, 
\int_{C_k^{(a)}} \,\mathrm d\bm s_k^{(a)}\,\int_{D_k^{(a)}} \,\mathrm d\bm\gamma_k^{(a)}\,
\sum_{\substack{c_1,\ldots,c_n=1: \\ \#\{c_1, \ldots, c_n\} = k}}^k\mathbb I[\bm c\in E_k^{(a)}]\\
&\hspace{5cm} \times \sum_{\ell=0}^\infty\frac{1}{\ell!}
\int_{B_\ell^{(na)}}\,\mathrm d\bm\mu_\ell^{(na)}\, 
\int_{C_\ell^{(na)}} \,\mathrm d\bm s_\ell^{(na)}\,\int_{D_\ell^{(na)}} \,\mathrm d\bm\gamma_\ell^{(na)}
\, \int_{F}\,\mathrm du.
\end{aligned}
\end{equation} 

Without introducing the ancillary variable $u$, that is, when leaving out the term $p(u\mid t)$ in \eqref{eq:post-new}), it becomes difficult to derive the full conditionals for the allocated and non-allocated variables $\bm\mu^{(a)},\bm s^{(a)},\bm\gamma^{(a)},\bm\mu^{(na)},\bm s^{(na)},\bm\gamma^{(na)}$. This is due to the term $1/t^n$ in \eqref{eq:post-new}, noting that $t = \sum_{h=1}^k s^{(a)}_h + \sum_{h=1}^\ell s^{(na)}_h$,
which makes it impossible to factorize according to 
 the allocated and non-allocated variables. When including $u$ we obtain that
\begin{equation}\label{eq:u}
p(u \mid t) \frac{1}{t^{n}} = \frac{u^{n-1}}{(n-1)!} \exp(-ut) t^{n} \frac{1}{t^{n}} = \frac{u^{n-1}}{(n-1)!} \left[\prod_{h=1}^k \exp(- u s^{(a)}_h)\right]\left[ \prod_{h=1}^\ell \exp(-u s^{(na)}_h)\right],
\end{equation}
which
does not depend on $t$. Using \eqref{eq:post-new} and \eqref{eq:u}, a factorization is obtained which is useful for the Metropolis-within-Gibbs sampler described in the following section. 

\section{Algorithm for posterior simulation}\label{sec:algorithm}

\subsection{Metropolis-within-Gibbs sampler}

 In our Metropolis-within-Gibbs sampler for simulating from the posterior \eqref{eq:post-new}, a single iteration is given by updating from full conditionals for five blocks of variables 
 as specified by the first line in the following steps (A)-(E).
  Note that we use the notation $p(\cdot|\cdots)$ to indicate that we consider a variable or collection of variables $\cdot$ given the remaining variables $\cdots$ (including the data $\bm y$).  
\begin{enumerate}
    \item[(A)] Update the non-allocated variables $(\bm \mu^{(na)}, \bm s^{(na)}, \bm \gamma^{(na)})$ from their full conditional as given by the following steps (i)-(iii), noting the following.
    Since the cardinality of each of the vectors $\bm s^{(na)}$ and $\bm \gamma^{(na)}$ agrees with the cardinality of $\bm \mu^{(na)}$,
    it is of paramount importance to resort to a \textit{collapsed} Gibbs sampler.
     Therefore, in (i) we sample $\bm \mu^{(na)}$ from the conditional density  obtained by integrating out $(\bm s^{(na)}, \bm \gamma^{(na)})$, and then
in (ii)-(iii) we sample $\bm s^{(na)}$ and $\bm \gamma^{(na)}$ from their respective full conditionals, hence knowing the cardinality $\ell$ of $\bm \mu^{(na)}$.   
    \begin{enumerate}
        \item[(i)] Sample from the conditional density obtained by integrating out $(\bm s^{(na)}, \bm \gamma^{(na)})$  and given by
        \begin{equation}\label{e:cond-density1}
            p(\bm \mu^{(na)} \mid \xi,\bm\mu^{(a)},\bm s^{(a)}, \bm\gamma^{(a)},\bm c, u, \bm y) \propto p(\bm \mu^{(a)} \cup \bm \mu^{(na)} \mid \xi) \psi(u)^\ell
         \end{equation}
         with respect to $\mathrm d\bm\mu^{(na)}$.
         Here, $\psi(u)$ denotes the Laplace transform of the density $p(s_h)$, and we can got rid of the last term $\psi(u)^\ell$, since
         $\ell$ is the cardinality of $\bm \mu^{(na)}$. In Section~\ref{sec:mu_na} we verify \eqref{e:cond-density1} and give  details for simulation from \eqref{e:cond-density1}.
         If, after this update, $\ell =0$, the following two items (ii) and (iii) are  skipped.
         
         \item[(ii)] Sample $\bm s^{(na)}$ from its full conditional,
         \[
             p(\bm s^{(na)} \mid \cdots) \propto \prod_{h=1}^\ell p(s^{(na)}_h) \exp(-u s^{(na)}_h).
         \]
         That is, sample independently $\ell$ values from the exponential tilting of the prior density. Depending on the specific choice of $p(s^{(na)}_h)$, this can be done exactly or requires a Metropolis-Hastings step.
         
         \item[(iii)] Sample $\bm \gamma^{(na)}$ from its full conditional,
         \[
             p(\bm \gamma^{(na)} \mid \cdots) \propto \prod_{h=1}^\ell p(\gamma^{(na)}_h).
         \]
         That is, sample independently $\ell$ values from the prior density $p(\gamma^{(na)}_h)$.
    \end{enumerate}

    \item[(B)] Update the allocated variables $(\bm \mu^{(a)}, \bm s^{(a)}, \bm \gamma^{(na)})$:
        \begin{enumerate}
            \item[(i)] Sample $\bm \mu^{(a)}$ from its full conditional,
            \[
                p(\bm\mu^{(a)} \mid \cdots)  \propto p(\bm \mu^{(a)} \cup \bm \mu^{(na)} \mid \xi) \prod_{h=1}^k \left[\prod_{i: c_i = h} k(y_i \mid (\mu^{(a)}_h,\gamma^{(a)}_h))\right],    
            \]
            where by \eqref{e:general} we can replace $p(\bm \mu^{(a)} \cup \bm \mu^{(na)} \mid \xi)$ by $g(\bm \mu^{(a)} \cup \bm \mu^{(na)} \mid \xi)$. 
            We do this by updating each of $\mu^{(a)}_h$ from
            \[
                p(\mu^{(a)}_h \mid \cdots) \propto g(\mu^{(a)}_h \cup \{\bm\mu^{(a)} \setminus \{\mu^{(a)}_h\}\} \cup \bm\mu^{(na)}) \prod_{i: c_i = h} k(y_i \mid (\mu^{(a)}_h,\gamma^{(a)}_h)).    
            \]   
            Appendix \change{Section~\ref{sec:proposal} discusses how to construct a proposal density for sampling from $p(\mu^{(a)}_h \mid \cdots)$ using a Metropolis-Hastings step.}
            
            \item[(ii)] Sample $\bm s^{(a)}$ from its full conditional,
            \[
    p(\bm s^{(a)} \mid \cdots) \propto \prod_{h=1}^k (s^{(a)}_h)^{n_h} e^{-u s^{(a)}_h} p(s^{(a)}_h).
\]
Here, the $s^{(a)}_h$'s are independent conditional to
everything else, so they can be updated individually using a Metropolis-Hastings step. 

        \item[(iii)] Sample $\bm \gamma^{(a)}$ from its full conditional,
        \[
            p(\bm \gamma^{(a)} \mid \cdots) \propto \prod_{h=1}^k p(\gamma^{(a)}_h) \prod_{i: c_i = h} k(y_i \mid (\mu^{(a)}_h,\gamma^{(a)}_h)).
        \]
            Unless $p(\gamma^{(a)}_h)$ and $k(y_i \mid (\mu^{(a)}_h,\gamma^{(a)}_h))$ are conjugate, we apply a Metropolis step for the $\gamma_h^{(a)}$'s.   
        \end{enumerate}
        Since in this step we have conditioned with respect to $\bm c$ too, $k$ denotes the number of clusters and is fixed. 

    \item[(C)] Sample each $c_i$ from its full conditional, which is a discrete distribution over ${1, \ldots, k + \ell}$ given by
    \begin{align*}
        p(c_i = h\mid\cdots) & \propto s^{(a)}_h k(y_i \mid (\mu^{(a)}_h, \gamma^{(a)}_h)), \qquad h=1, \ldots, k, \\
        p(c_i = k + h\mid\cdots) & \propto s^{(na)}_h k(y_i \mid (\mu^{(na)}_h, \gamma^{(na)}_h)), \qquad h=1, \ldots, \ell.
    \end{align*}
    After this, with a positive probability it may happen that $c_i > k$ for some $i$'s, so that some non-allocated components have become allocated, and some allocated components have become non-allocated.
    Then a simple relabelling of $(\bm\mu^{(a)},\bm s^{(a)}, \bm\gamma^{(a)},\bm\mu^{(na)},\bm s^{(na)},\bm\gamma^{(na)})$ and $\bm c$ is needed, so that $\bm c$  takes values in $\{1, \ldots, k\}^n$.
    
    \item[(D)] Sample $\xi$ from its full conditional,
    \[
        p(\xi \mid \cdots) \propto p(\xi) p(\bm \mu^{(a)} \cup \bm \mu^{(na)} \mid \xi).
    \]
    This requires a Metropolis-Hastings step, which is not straightforward when $Z_\xi$ in \eqref{e:general} is not expressible in closed form, e.g.\ in the case of a repulsive pairwise interaction point process. Details on how this issue is overcome are given in Section~\ref{sec:xi}.
    
    \item[(E)] Sample $u$ from its full conditional, which is just a gamma distribution with shape parameter $n$ and inverse scale $t$.
\end{enumerate}

\subsection{Updating the non-allocated variables}\label{sec:mu_na}

This section provides the remaining details of step (A)(i). 
By \eqref{eq:post-new},
\begin{align}
     p(\bm \mu^{(na)} \mid \xi,&\bm \mu^{(a)}, \bm S^{(a)}, \bm \gamma^{(a)}, \bm c, u, \bm y) \nonumber\\ 
     &= \int \int p(\bm \mu^{(na)}, \bm s^{(na)}, \bm \gamma^{(na)} \mid \xi,\bm \mu^{(a)}, \bm S^{(a)}, \bm \gamma^{(a)}, \bm c, u, \bm y) \, \mathrm{d}\bm s^{(na)} \, \mathrm{d}\bm \gamma^{(na)} \nonumber\\
     & \propto \int \int p(\bm\mu^{(a)}\cup\bm\mu^{(na)}\,|\, \xi)
 \left[\prod_{h=1}^\ell p(\gamma^{(na)}_h) p(s^{(na)}_h)\right] p(u \mid t) \frac{1}{t^n} \, \mathrm{d}\bm s^{(na)} \, \mathrm{d}\bm \gamma^{(na)} \nonumber\\
 & \propto \int p(\bm\mu^{(a)}\cup\bm\mu^{(na)}\,|\, \xi)
 \left[\prod_{h=1}^\ell \exp(-u s^{(na)}_h) p(s^{(na)}_h)\right] \, \mathrm{d}\bm s^{(na)} \label{e:aaa}\\
 & = p(\bm\mu^{(a)}\cup\bm\mu^{(na)}\,|\, \xi) \psi(u)^\ell\label{e:bbb}
\end{align}
where \eqref{e:aaa} follows by integrating over $\gamma^{(na)}_h$ and using \eqref{eq:u}, and \eqref{e:bbb} by applying the definition of $\psi(u)$. This verifies \eqref{e:cond-density1}.

Note that \eqref{e:cond-density1} identifies an unnormalized density for $\bm\mu^{(na)}$ with respect to  $\mathrm d\bm\mu^{(na)}$. 
In our examples, the unnormalized density \change{in \eqref{e:bbb} will be hereditary, that is, $p(\bm\mu^{(a)}\cup\bm\mu^{(na)}\,|\, \xi)>0$ implies $p(\bm\mu^{(a)}\cup\bm\mu'^{(na)}\,|\, \xi)>0$ whenever $\bm\mu'^{(na)}$ consists of one more point than $\bm\mu^{(na)}$. Moreover, in our examples, this density is
defined on a compact set, and so we can easily} employ 
the birth-death Metropolis-Hastings algorithm in \cite{geyer1994simulation}. Specifically, we use
Algorithm 11.3 in \cite{moller2003statistical}.

\subsection{Sampling the hyperparameter $\xi$}\label{sec:xi}

When the density $p(\bm \mu \mid \xi)$ is expressible in close form, a standard Metropolis-Hastings move can be employed to update $\xi$ from its full conditional.
However, when $Z_\xi$ is intractable, it is a doubly-intractable problem, since a ratio of unknown normalizing constants appears in the Hastings ratio.
In fact, if $p(\xi^\prime ; \xi\mid\cdots)$ is a proposal density for the Metropolis-Hastings step for the full conditional of $\xi$, then the acceptance ratio amounts to
\begin{equation}\label{e:acc-jm}
    \alpha(\xi^\prime; \xi\mid\cdots) = \frac{p(\xi^\prime) g(\bm \mu \mid \xi^\prime) p(\xi; \xi^\prime)}{p(\xi) g(\bm \mu \mid \xi) p(\xi^\prime ; \xi)} \frac{Z_\xi}{Z_{\xi^\prime}},
\end{equation}
which is intractable due to the term $Z_\xi / Z_{\xi^\prime}$.
To overcome this issue, we can use the exchange algorithm described in 
\cite{murray2006mcmc} and inspired by the \textit{single auxiliary variable method}
proposed by \cite{moller2006efficient}. \change{For further details, see Appendix~\ref{sec:exchange}.}
This algorithm requires generating an ancillary variable following the same distribution of $\bm \mu$ given $\xi^\prime$.
To this end, we employ the dominated coupling from the past algorithm in \cite{kendall2000perfect}.

In previous literature, the use of the ancillary variable algorithms in \cite{moller2006efficient} and \cite{murray2006mcmc} has been limited because of the high computational cost associated to perfect simulation.
In contrast, in our examples perfect simulation is fast.
As an example, approximating the density of a DPP with  $N=50$ in dimension $q=2$, as done in \cite{bianchini2018determinantal}, is around $25$ times more
expensive than running a perfect simulation from a Strauss process with parameters
and prior for $\xi$ chosen as in Section~\ref{sec:strauss_params}; \change{for further comparisons,
see Appendix~\ref{sec:strauss_vs_dpp}.}
The perfect simulation step is very fast  since $m$ (the number of components in the mixture model) is typically small (in our examples it is  always less than 10).
However, when dealing with applications with a very large number of clusters, such as topic modeling, where the number of clusters is usually between 50 and 100, cf.\ \cite{blei2003latent}, we expect
perfect simulation to be potentially a bottleneck and
limit the use of the exchange algorithm.
\change{Although not investigated here, in such cases we could avoid perfect simulation by replacing the exact exchange algorithm of \cite{murray2006mcmc} with asymptotically exact algorithms that should offer smaller computational cost; see for instance \cite{lyne2015russian} and \cite{liang2016adaptive}.}

\section{Prior elicitation}\label{sec:prior}

In this section, we discuss prior elicitation and how to set hyperparameters when the
prior for $\bm\mu$ is the Strauss process or the DPP with power exponential spectral density.

\subsection{Prior elicitation for the Strauss process}\label{sec:strauss_params}

Consider the Strauss process prior given by \eqref{e:pairwise}-\eqref{eq:DPP_phi2}.
In addition to the parameter $\xi$ which controls the intensity, the process depends on two parameters $\alpha \in[0,1]$ and $\delta > 0$ which  control repulsiveness and the range of interaction, respectively.  
Initially we investigated cases where $\alpha$ and $\delta$ were random, but then our simulated datasets  yielded a large number of clusters a posteriori. 
Moreover, when fitting mixtures of Gaussian densities to data generated from heavy-tailed distributions, 
as also discussed at the beginning of Section~\ref{sec:simulation}, 
in general better density estimates 
were obtained when using a larger (i.e., larger than the true value) number of components in the mixture model.
For this reason, we obtained  a posteriori values of $\alpha$ and $\delta$ that induced less repulsive behaviors than  desired. 
Therefore, we suggest below to fix $\alpha$ and $\delta$ via an empirical Bayes procedure, 
and let only $\xi$ be random.

We propose to estimate $\alpha$ and $\delta$ as follows. Denote by $p(r)$ the kernel density estimate of the empirical distribution of the pairwise distances between  observations; in all the examples, we have obtained such an estimate using the default bandwith selection procedure in Python's \texttt{scipy} package.
Since
 $\delta$ should be large enough to induce repulsion of redundant clusters, but not too large to affect density estimation severely,
  we suggest to fix $\delta$ as the smallest local minimum point of $p(r)$, that is, $\delta = \min_{r > 0} \{r : \ r \text{ is a local minimum for } p(r)\}$.
Further, $\alpha$ should be small enough to encourage separation between the allocated means.
Consider, for example, the case with two clusters $\{y_i:\,c_i=h'\}$ and $\{y_i:\,c_i=h''\}$ where $0<\|\mu_{h'}-\mu_{h''}\|\le\delta$ but the distances from $\mu_{h'}$ and $\mu_{h''}$ to all the other $\mu_h$'s are greater than $\delta$.
Then, by \eqref{eq:post}, the full conditional of $\mu_{h'}$ has density
\[
    p(\mu_{h'} \mid \cdots) \propto \alpha\prod_{i:\,c_i=h'}k(y_i\mid\mu_{h'},\gamma_{h'}).
\]
Now, the point of using a repulsive prior is that if the cardinality of cluster $h'$, that is $\#\{i:\,c_i=h'\}$, is small, the repulsiveness should prevail on the within-cluster likelihood:
That is, regardless of how well the value of parameter $\mu_{h'}$ \virgolette{fits} data in cluster $h'$, the full conditional density associated to that value should be small because $\mu_{h'}$ is near to the cluster center $\mu_{h''}$.
A rough estimate of $\alpha$ can be obtained 
by assuming $\alpha = \exp(-n^* \log(k_s))$. 
Here, $n^*$ represents the minimum cluster size needed to balance the repulsive behavior induced by the prior, while $k_s$ represents a \virgolette{guess} of $k(\cdot \mid \cdot)$ in a small cluster.
In our experiments, we assumed that clusters with less than $5\%$ of the data should be considered small and thus we fixed $n^* = n / 20$.
Further, we fixed $\log(k_s) = 1$ so that this term did not affect the definition of $\alpha$. In addition,
preliminary sensitivity analysis on $\alpha$ led us to conclude that posterior inference is robust.

Finally, we assume that $\xi$ is random. An upper bound for the expected number of points in $\bm\mu$ is $\xi  |R|$, and given an upper bound $M_{\max}$ on the expected number of components, we assume the prior for $\xi$ to be uniform over the interval
$\left(|R|^{-1}, M_{\max} |R|^{-1} \right)$.
Since the number of clusters is smaller than the number of components, 
$M_{\max}$ is an upper bound for both, to be fixed in each application according to prior belief.

\subsection{Prior elicitation for the power exponential spectral DPP model}\label{sec:dpp_params}

For the DPP defined on $\mathbb R^q$ by the spectral density $\chi$ used in \eqref{eq:spectral}, existence is ensured if
$0<\alpha \le \alpha_{\max}$, where
\[
    (\alpha_{\max})^q = \frac{\pi^{q/2} \Gamma(q/\beta + 1)}{\xi \Gamma(q/2 + 1)},
\]
cf.\ \cite{lavancier2015determinantal}. 
So we let $\alpha = s \ \alpha_{\max}$ with $0 < s < 1$ (as specified below), which implies existence of the DPP restricted to any compact subset  of $\mathbb R^q$. 
Note that the DPP density given by \eqref{e:C}-\eqref{eq:spectral} refers to the case $R=[-1/2, 1/2]^q$, and a simple
rescaling is needed in the density expression when we fix $R$ to be the smallest rectangle containing all the observations, cf.\ \cite{lavancier2015determinantal}.

Recall that $\xi$ is the expected number of points in $\bm\mu$. 
We let a priori $\xi$ be 
uniformly distributed over $[1, M_{\max}]$, 
where $M_{\max}$ is fixed (as in the case of the Strauss process, cf.\ Section~\ref{sec:strauss_params}).
As noted in \cite{lavancier2015determinantal}, the parameters $(s, \beta)$ are harder to interpret  via \eqref{eq:spectral}.
In our examples, we fix $s=0.5$ and perform sensitivity analysis on $\beta$, concluding that inference is robust.

\section{Simulation studies}\label{sec:simulation}

In this section, we compare the reversible jump algorithm in \cite{bianchini2018determinantal}  to our Metropolis-within-Gibbs sampler
presented in Section~\ref{sec:algorithm}, and  show the advantages of repulsive mixtures over non-repulsive ones.
We refer to our Metropolis-within-Gibbs sampler as the \virgolette{M-w-G sampler} and the reversible jump algorithm as \virgolette{RJ}.
\change{In Appendix~\ref{sec:extra_sim}}, we illustrate the advantages of using a Strauss
process over a DPP as prior for $\bm \mu$ \change{and provide further simulations when the dimension $q$ or the number of components $m$ increase.}
In particular, we conclude that the computational cost of posterior inference under the DPP grows exponentially with data dimension $q$, whilst the computational cost associated to the Strauss process is almost constant as data dimension increases, and that posterior summaries obtained under the DPP and Strauss process are almost identical.  This motivates the use of the Strauss process as  a prior for $\bm \mu$.

In this section, we study posterior inference in \textit{misspecified} settings, i.e., when the generating process does not coincide with the model used to fit data; 
for a formal definition of misspecification, see \cite{kleijn2006misspecification}.
In misspecified settings, there is a trade-off between the accuracy of the density and  number of clusters estimation recovered by the mixture model, cf.\ \cite{guha2019posterior}. 
This indicates that more accurate density estimates correspond to overestimated number of clusters and vice-versa. In fact,  to recover the shape of non-Gaussian data, several Gaussian components (with similar values of the mean parameters) are needed.   
We expect that the repulsiveness induced by the prior for $\bm \mu$ favours cluster  over density estimation.

We consider two simulation scenarios, 
the first one is as in \cite{miller2019robust}, where the  authors  
generated iid data $y_1,\ldots,y_n$ using a two-step procedure as follows.
First, a mixture density $f_0$  with $m_0$ components is selected. Second, a random
density $\widetilde f$ is drawn from a Dirichlet process mixture, with base measure given by $f_0$.
Specifically,
\begin{equation}\label{eq:dgp1}
\begin{aligned}
    y_1, \ldots, y_n \mid P &\iid \widetilde f(\cdot) = \int \calN(\cdot \mid \theta, 0.25^2) P(\mathrm{d} \theta), \\
     P \sim DP(a f_0), & \qquad f_0 = \sum_{h=1}^{m_0} w_{0h} \, \calN(\mu_{0h}, \sigma^2_{0h}),
\end{aligned}
\end{equation}
where $DP(a f_0)$ denotes the Dirichlet process with total mass parameter $a$ and centering probability measure induced by $f_0$. 
We fix $a = 500$, $m_0 = 4$, $\bm w_0 = (0.25, 0.25, 0.3, 0.2)$, 
$\bm \mu_0 = (-3.5, 3, 0, 6)$,
and $\bm \sigma_0 = (0.8, 0.5, 0.4, 0.5)$. 
Following \cite{miller2019robust}, we interpret the data generating density $\widetilde{f}$
as a perturbation of the \virgolette{true} density $f_0$,  so the goal is
to recover $f_0$ and $m_0$.

The second  simulation scenario considers draws from the following mixture of two components:
\begin{equation}\label{eq:dgp2}
    y_1, \ldots y_n \iid 0.5 \, t_q(1, \bm \mu_0, \Sigma_0) + 0.5 \, MSN_q(\omega, \mu_1, \sigma_1) .
\end{equation}
Here $t_q(1, \bm \mu_0, \Sigma_0)$ denotes the density of a $q$-dimensional Student distribution with one degree of freedom, location $\bm \mu_0$, and scale matrix
$\Sigma_0$. Furthermore, $MSN_d(\omega, \mu_1, \sigma_1)$ denotes the density of a $q$-dimensional random vector, where each entry is drawn independently from a skew normal distribution  with
mean $\mu_1 + \omega \sigma_1 \sqrt{2 / \pi}$,  being $\mu_1$ the location parameter and $\omega$ the scale parameter of the skew normal distribution. The dimension $q$ and the other parameters in \eqref{eq:dgp2} will be specified later. 

For both scenarios, the kernel $k(\cdot \mid \cdot)$ in \eqref{eq:mixture}
is either the univariate or the multivariate Gaussian density.
In addition to the prior assumptions (a)-(e) in Section~\ref{sec:1.1}, we let a priori 
$(w_1, \ldots, w_m) = (s_1 /t, \ldots, s_m/t)$, with $\bm s$ as in Section~\ref{sec:prior_weights}, where each $s_h$ follows a gamma distribution with shape and  scale equal to one,
 and assume each $\gamma_h$ to be either inverse-gamma distributed (if data are univariate) or inverse-Wishart distribution (for multivariate data), with hyperparameters fixed as in \cite{fraley2007bayesian}.
Finally, unless otherwise stated,  parameters of the Strauss point process or the DPP are chosen as discussed in Sections \ref{sec:strauss_params}-\ref{sec:dpp_params}.
In particular, we fix $M_{\max} = 30$.

\subsection{Monitoring MCMC mixing}
\label{sec:vs_rj}

\begin{table}[t]
\centering
\begin{tabular}{ | c | c || c | c | c | c | c | }
    \hline
    \multicolumn{2}{|c||}{Params.} & \multicolumn{2}{ |c |}{RJ} & \multicolumn{3}{|c|}{M-w-G sampler} \\
    \hline
    $\xi$ & $\beta$ & $ESS$ & $\mathbb E[m \mid \mbox{data}]$ & $ESS$ & $\mathbb E[m \mid \mbox{data}]$ & $\mathbb E[k \mid \mbox{data}]$ \\
    \hline
    4 & 10 &  90.63 & 4.33 & 8201.41 & 4.01 & 4.00 \\
    4 & 2.5 &  62.46 & 4.402 & 3735.80 & 4.01 & 4.00 \\
    4 & 25 &  83.32 & 4.44 & 2971.05 & 4.02 & 4.00 \\
    \hline
\end{tabular}
\caption{Summary of the MCMC simulations for the reversible jump algorithm (RJ) in \cite{bianchini2018determinantal} and our Metropolis-within-Gibbs sampler (M-w-G sampler). ESS denotes the effective sample size out of the $10,000$ MCMC samples.}
\label{tab:vs_rj}
\end{table}

In this section, data are given by 500 observations simulated in accordance to 
\eqref{eq:dgp1}. The marginal prior for $\bm \mu$ is the DPP 
specified in Section~\ref{sec:dpp_params},  where in order to identify the effect of the algorithm on posterior inference, we  keep the intensity
parameter $\xi$ fixed.
Furthermore, we consider three possible values for the hyperparameters  $\xi$ and $\beta$ in the DPP prior, cf.\ Table~\ref{tab:vs_rj}.
For each  choice of hyperparameter values, 
we ran both MCMC algorithms (M-w-G sampler and RJ) for $20,000$ iterations, discarding the first $10,000$ as burn-in and without thinning the chain.
In order to compare the results, we  consider the effective sample size (ESS)
of the number of components in the mixture ($m$ in our notation) as well
as its autocorrelation.

\begin{figure}[t]
    \centering
    \begin{subfigure}{0.5\linewidth}
        \centering
        \includegraphics[width=\linewidth]{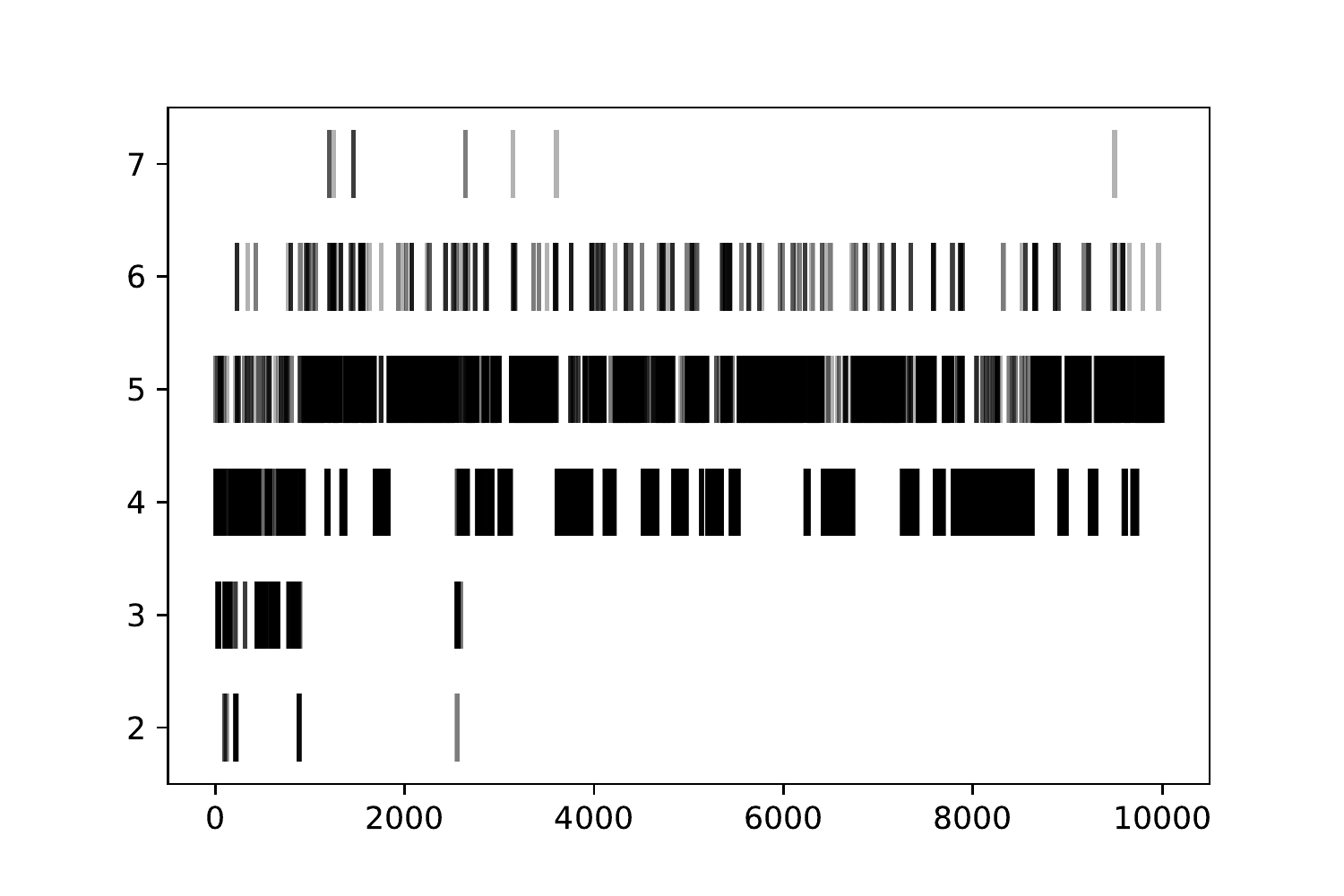}
    \end{subfigure}%
    \begin{subfigure}{0.5\linewidth}
        \centering
        \includegraphics[width=\linewidth]{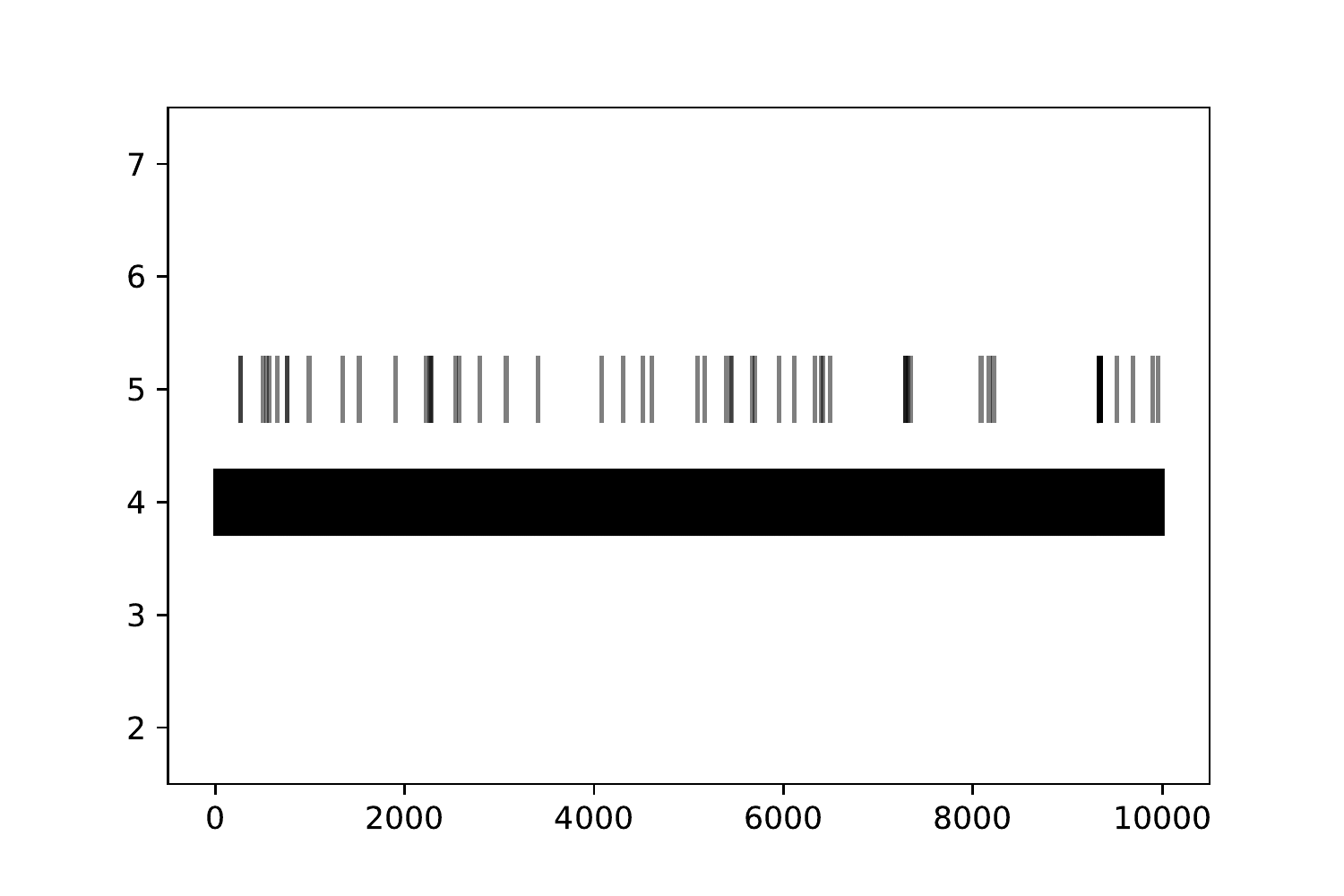}
    \end{subfigure}
    \begin{subfigure}{0.5\linewidth}
        \centering
        \includegraphics[width=0.8\linewidth]{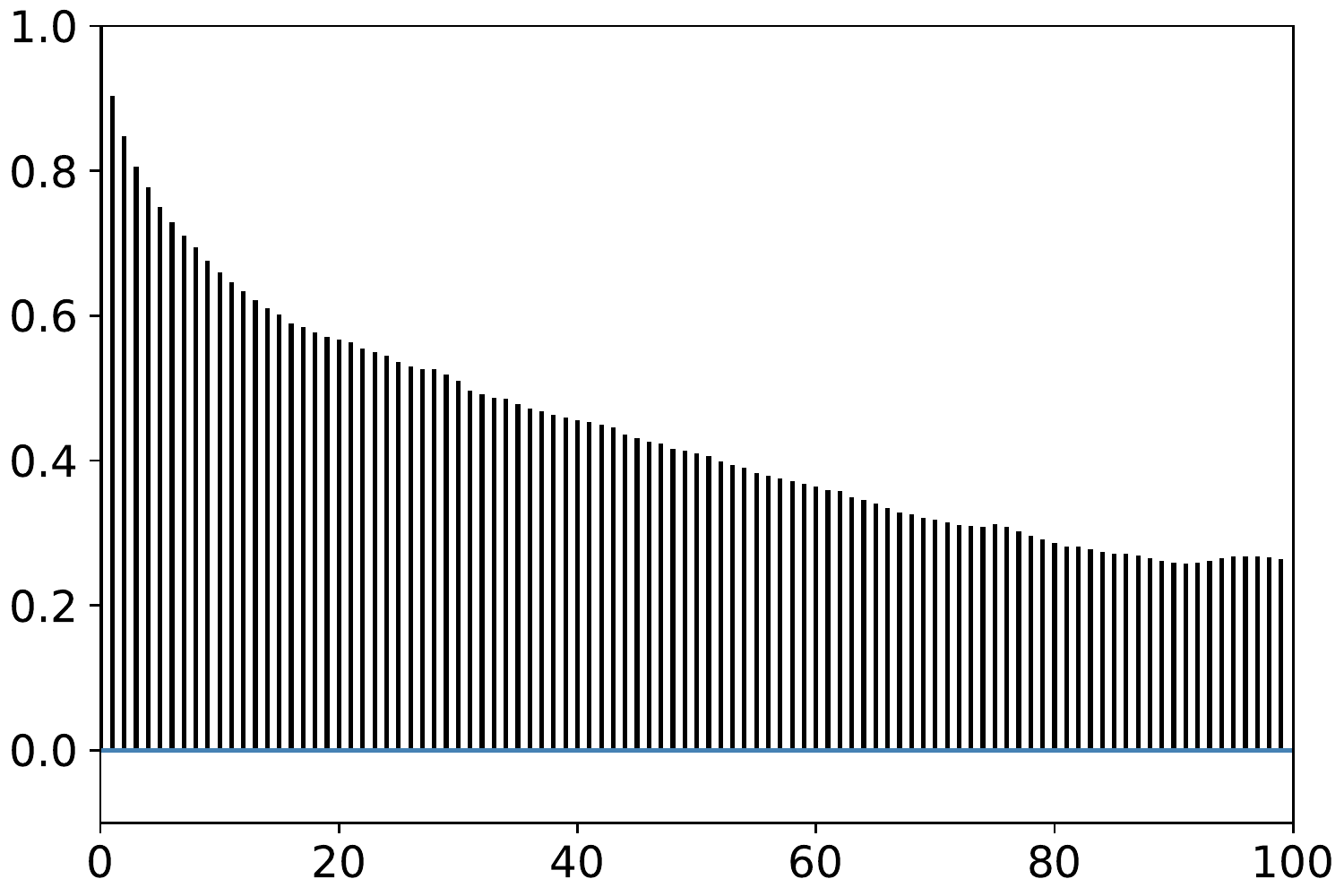}
    \end{subfigure}%
    \begin{subfigure}{0.5\linewidth}
        \centering
        \includegraphics[width=0.8\linewidth]{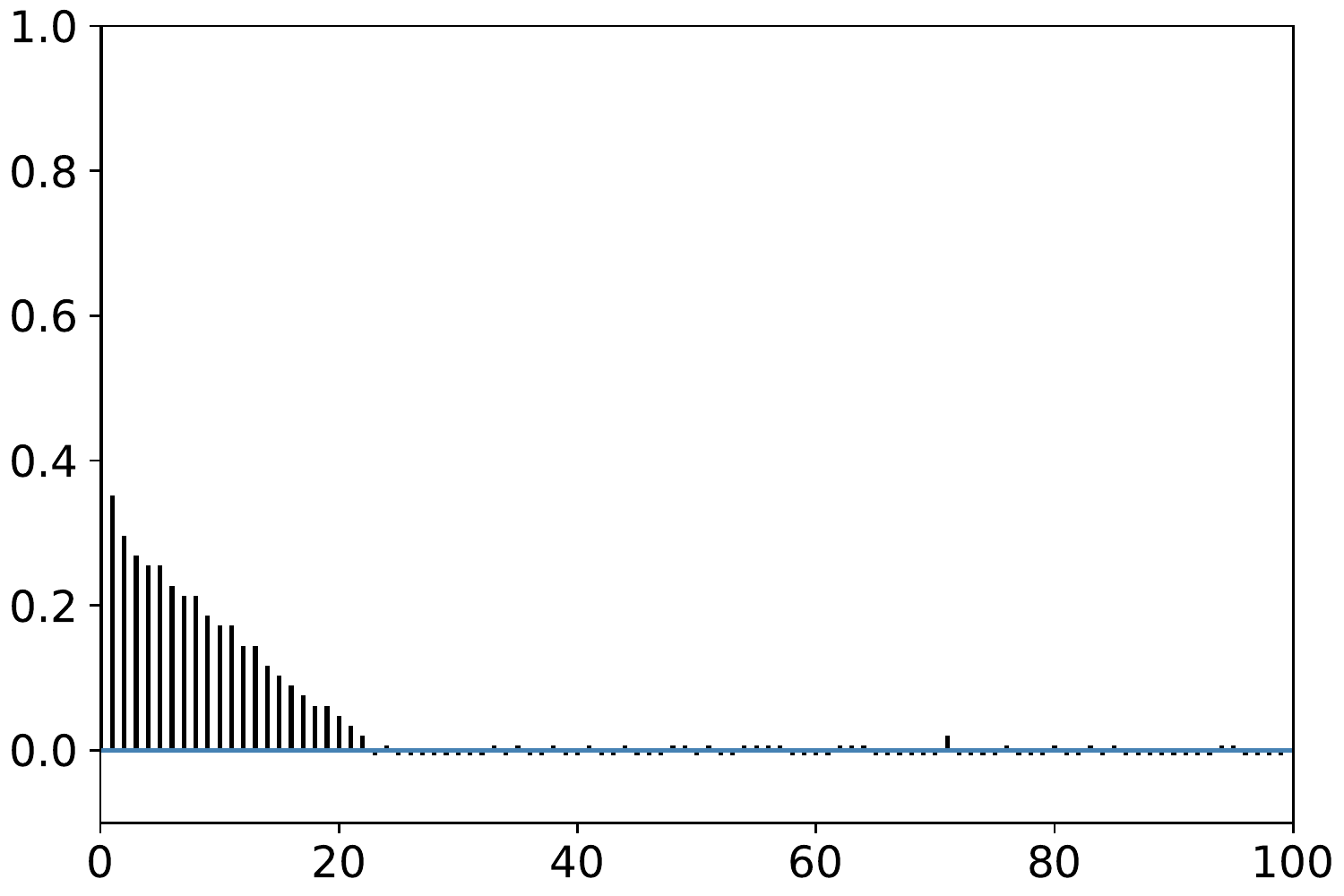}
    \end{subfigure}
    
    \caption{Trace plots (top) and autocorrelations (bottom) of $ m$ when $\xi = 4$ and $\beta=10$. Left: RJ. Right: M-w-G sampler.}  
    \label{fig:vs_rj}
\end{figure}

Table~\ref{tab:vs_rj} reports, for different combination of hyperparameters, the posterior expected value of $m$ as well
as the effective sample sizes for $m$ obtained
by the two algorithms.
Since in our M-w-G sampler the number of clusters can be smaller
than $m$, the table also shows the posterior expected value of $k$ (the number of allocated components/clusters). 
Figure~\ref{fig:vs_rj} shows for both algorithms trace plots  and autocorrelation plots for $m$ when $\xi = 4$ and $\beta=10$ (first row of Table~\ref{tab:vs_rj}). Note that both algorithms offer
good estimates of the number of components in 
the mixture.
However, the performance
of our M-w-G sampler is superior to the RJ algorithm
in all the settings of hyperparameters we tested:
Our \change{M-w-G sampler} generally produces a (much) higher effective sample size
and overall better mixing of the chains.

\subsection{Comparison with DPM and FM}\label{sec:vs_dpm_mfmf}

We focus on the differences between repulsive and non-repulsive mixtures using two further simulations.
For the class of non-repulsive mixtures, we consider (i) the finite mixture models (FM) of Gaussian densities in \cite{argiento2019infinity} and \cite{miller2018mixture}, and (ii) the Dirichlet Process Mixture (DPM) of Gaussian densities.

Both FM and DPM require the choice of a base measure $P_0$ that we fix as the normal-inverse-Wishart distribution (or the normal-inverse-gamma distribution in the univariate case).
Hyperparameters are  fixed according to \cite{fraley2007bayesian} to provide 
a weakly informative prior. Moreover, the concentration parameter in the Dirichlet process is fixed to one, and for the FM model we consider as prior for $m$ the shifted Poisson distribution (with support $\{1, 2, \ldots \}$)  so that the
 prior mean of the components  is equal to four if the data generating process is \eqref{eq:dgp1} and  to two if the data generating process is \eqref{eq:dgp2}.  Finally, for
 our model, we assume the Strauss process prior for $\bm \mu$.

Posterior simulation from the FM model was carried out using the R package \texttt{AntMAN}\footnote{available at https://github.com/bbodin/AntMAN}, while for the DPM we used the R package \texttt{BNPMix} \citep{bnpmix}.
For all three models, we ran the MCMC algorithm for $100,000$ iterations, discarding the first $50,000$ as a burn-in and keeping one of every ten iterations, so that in each case the final sample size is $5,000$.

\begin{figure}[t]
\begin{subfigure}{0.5\linewidth}
    \centering
    \includegraphics[width=\linewidth]{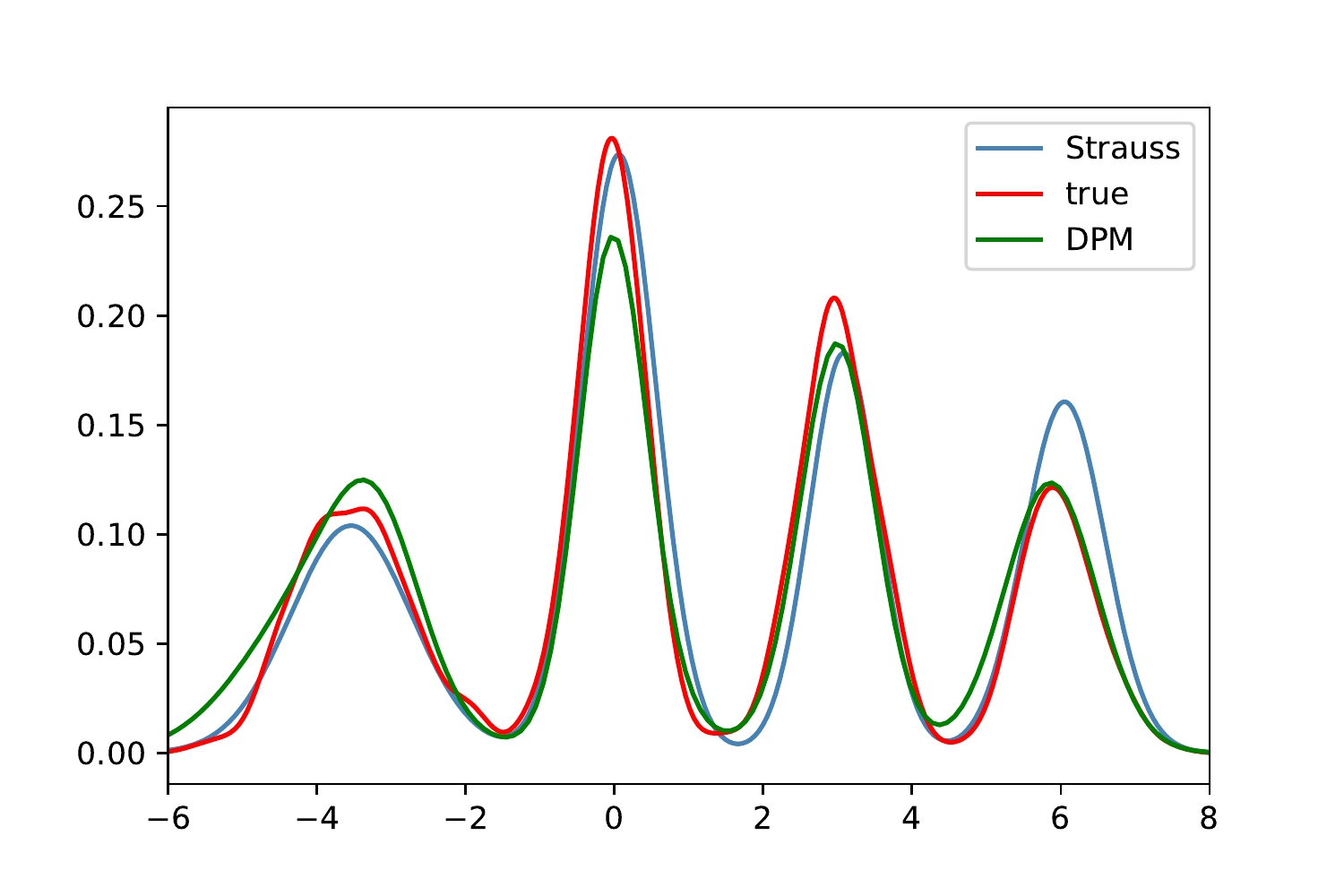}
\end{subfigure} %
\begin{subfigure}{0.5\linewidth}
    \centering
    \includegraphics[width=\linewidth]{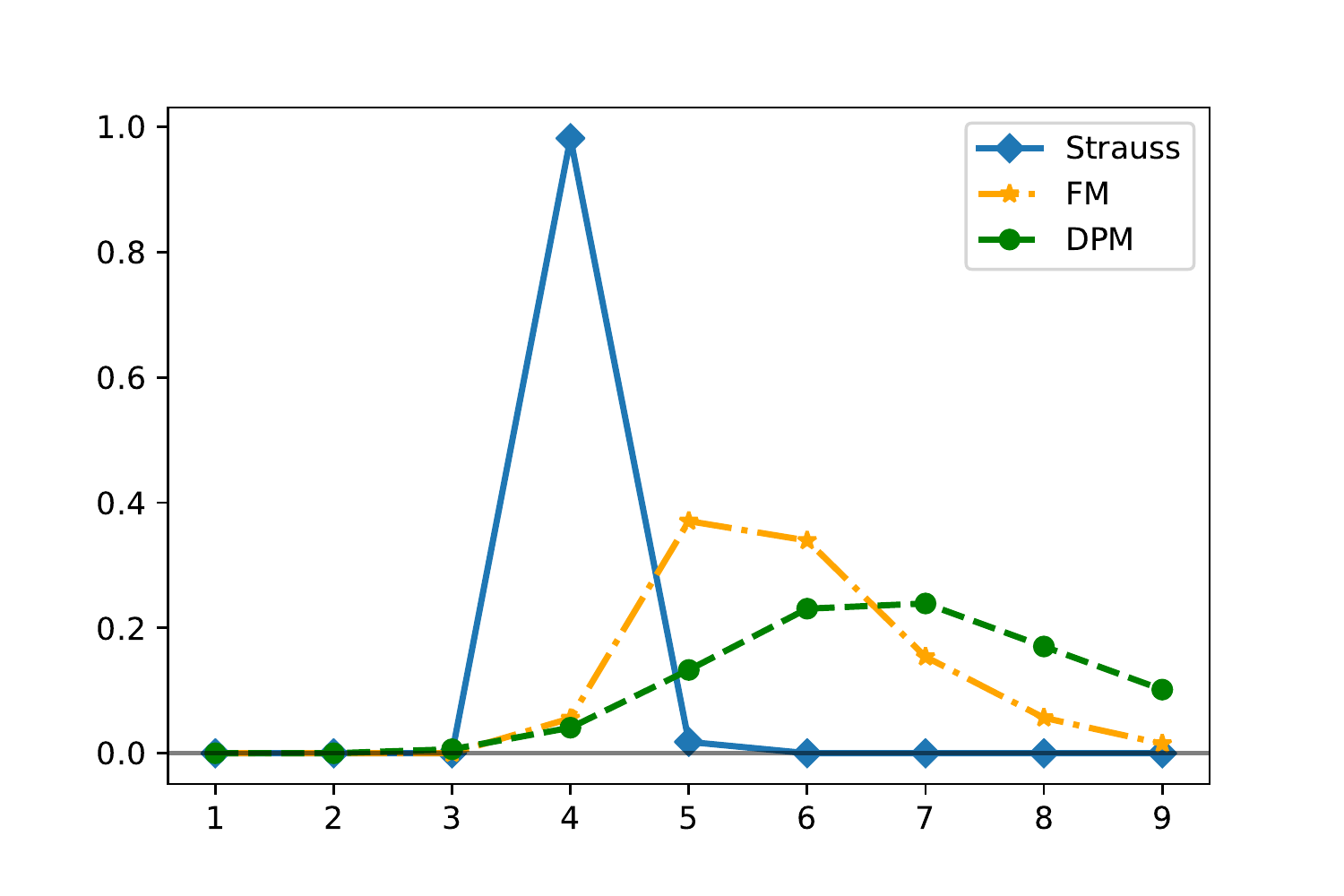}
\end{subfigure}
\caption{Posterior inference based on data simulated from \eqref{eq:dgp1}. To the left,  Bayesian mixture density estimates under the Strauss process and the DPM priors for $\bm\mu$, together with the true mixture density which has four components. To the right, posterior distributions of the number of allocated components under the Strauss process, FM, and DPM  priors for $\bm\mu$.}
\label{fig:sim_perturbed}
\end{figure}

In the first simulation study, data are given by 400 simulated observations from \eqref{eq:dgp1}.
Figure~\ref{fig:sim_perturbed} shows the true data generating density,  together with Bayesian mixture density estimates obtained by our model and the DPM (left),
as well as  the distribution of the number of clusters under the three
models (right). Here, by Bayesian density estimate we always mean the posterior expectation of the mixture density evaluated on a fixed grid of
points.
As expected, under this misspecified setting,  the use of non-repulsive mixture models
overestimate the number of clusters.
For instance, to recover the shape of the leftmost \textit{bell} of the data generating density in Figure~\ref{fig:sim_perturbed}, several Gaussian components (with  close cluster centers) are needed. 
Our model instead,  due to the repulsiveness induced by the prior on $\bm \mu$, \virgolette{correctly} identifies four clusters.

\begin{figure}
    \begin{subfigure}{0.45\linewidth}
    \centering
    \includegraphics[width=\linewidth]{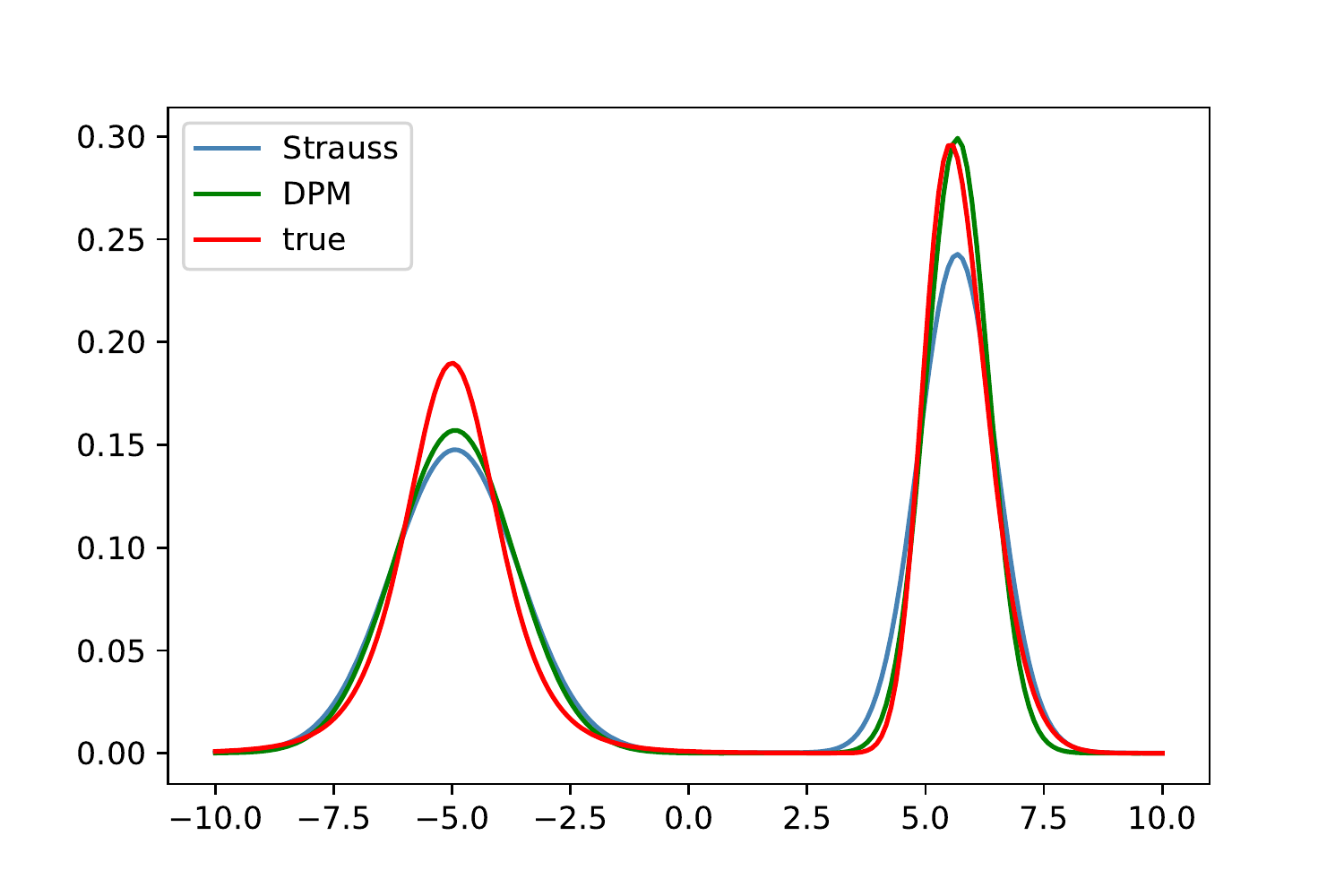}
\end{subfigure} %
\begin{subfigure}{0.6\linewidth}
    \centering
    \includegraphics[width=\linewidth]{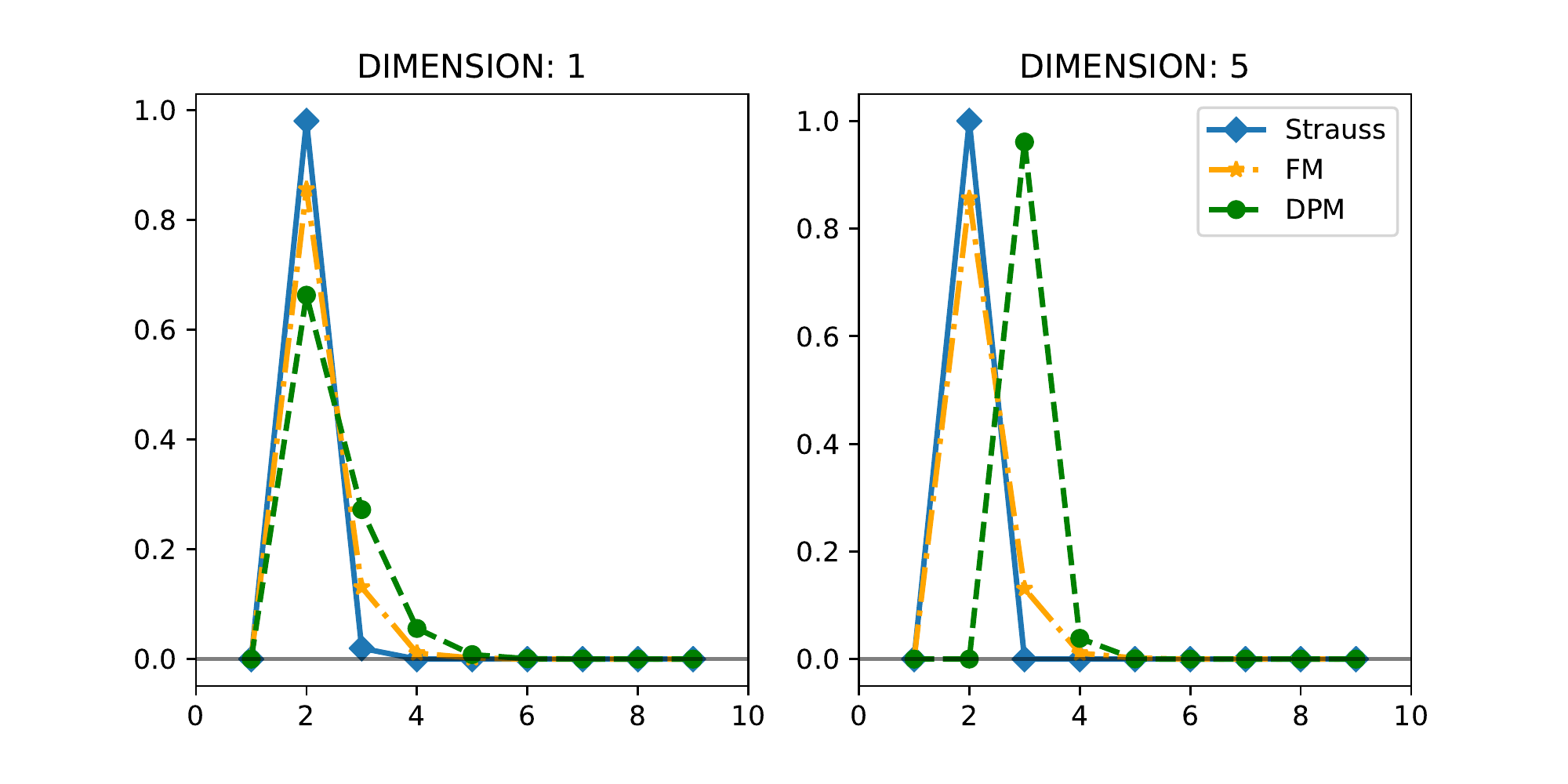}
\end{subfigure}
\caption{Posterior inference based on data simulated from \eqref{eq:dgp2}. To the left, when $q=1$,  Bayesian
mixture density estimates under the Strauss process and the DPM priors for $\bm\mu$, together with the true mixture density which has two components.
To the right, when $q=1,5$, posterior distributions of the number of allocated components under the Strauss process, FM, and DPM  priors for $\bm\mu$.}
\label{fig:sim_t_skew}
\end{figure}

For the second simulation study, we simulated 500 observations from \eqref{eq:dgp2} in each of the cases $q=1$ and $q=5$, where
 we fixed $\mu_0 = (-5, \ldots, -5)$ , $\Sigma_0 = I_q$,  $\omega = 2$, $\mu_1 = 5$, and $\sigma_1 = 1$.
Figure~\ref{fig:sim_t_skew} reports density estimates when $q=1$ (left) and the posterior distribution of the number of clusters for the three models (right) when $q=1,5$.
Note that, among the three models, our is the one that gives highest posterior probability to the true value $k=2$. When $q=5$, DPM assigns the highest probability to  three clusters. 
\change{Appendix~\ref{sec:clust} contains a comparison of the cluster estimates under the three models considered when $q=1$, and we conclude that the repulsive mixture model is the one that better recovers the true clustering of the data in this example.}

\section{Teenager problematic behavior dataset}\label{sec:teenage}

In this section,
we apply our model, with the Strauss process prior for $\bm\mu$, to a dataset consisting of 
$n=6504$ observations coming from the  %
Wave 1 data of the National Longitudinal Study of Adolescent to Adult Health,  which
 is available at \url{http://www.icpsr.umich.edu/icpsrweb/ICPSR/studies/21600}.  The data were also considered in \cite{collins2009latent} and \cite{li2018bayesian}. 

The dataset corresponds to six survey items pertaining problematic 
behaviors in teenagers, so that for the $i$'th teenager, $y_i=(y_{i1},\ldots,y_{i6}) \in \{0, 1\}^6$ is a binary vector, where $y_{ij}=1$ means a positive answer to entry $j$.
The  six entries correspond to
\begin{enumerate*}[label=(\roman*)]
    \item `lied to parents',
    \item `loud/rowdy/unruly in a public place',
    \item `damaged property',
    \item `stolen from a store',
    \item `stolen something worth less than $50$ dollars', and
    \item `taken part in a group fight'.
\end{enumerate*}

We let the kernel in \eqref{eq:mixture} be given by
\begin{equation}\label{eq:kernel_latent_class}
    k( y \mid \mu_h) = \prod_{j=1}^6 \mu_{hj}^{y_j} \ (1 - \mu_{hj})^{1 - y_j}, \quad y=(y_1,\ldots,y_6) \in\{0,1\}^6,
\end{equation}
so that the six entries in $ y_i=(y_{i1},\ldots,y_{i6})$ are conditionally independent binary  random
variables with success probability vector $ \mu_h=(\mu_{h1},\ldots,\mu_{h6})$.
Note that there is no parameter $\gamma$,  and the probability vector 
$(\mu_1, \ldots, \mu_m)$ belongs to $R = [0, 1]^6$.
The mixture model with kernel \eqref{eq:kernel_latent_class}
is known as a \textit{latent class model}. 

As the prior for $\bm\mu$, we assume the Strauss process on $R$ with parameters $\delta = 0.4$,
$\alpha =  \e^{-n^*}$ (with $n^* = 50$), and a uniform prior on $\xi$
with $M_{\max} = 30$, cf.\ Section~\ref{sec:strauss_params}.
In this context, we may consider the $\mu_h$'s as cluster centres/locations, where repulsion among the $\mu_h$'s is meant to favor identification of the clusters.

We ran our posterior simulation algorithm for $20,000$ iterations, after discarding other $20,000$
iterations as burn-in and saving one of every ten iterations. So the final sample is of
size $M=2,000$, and we denote $\mu_h^j$ the value of $\mu_h$ at iteration $j=1,\ldots,M$.
Below
we summarize our findings for the cluster centres and compare to what was obtained in 
\cite{li2018bayesian}, where the authors used a finite mixture model with
the same kernel \eqref{eq:kernel_latent_class} as ours, but fixed the number of clusters to be equal to four.

\begin{figure}[t]
\centering
    \includegraphics[width=\linewidth]{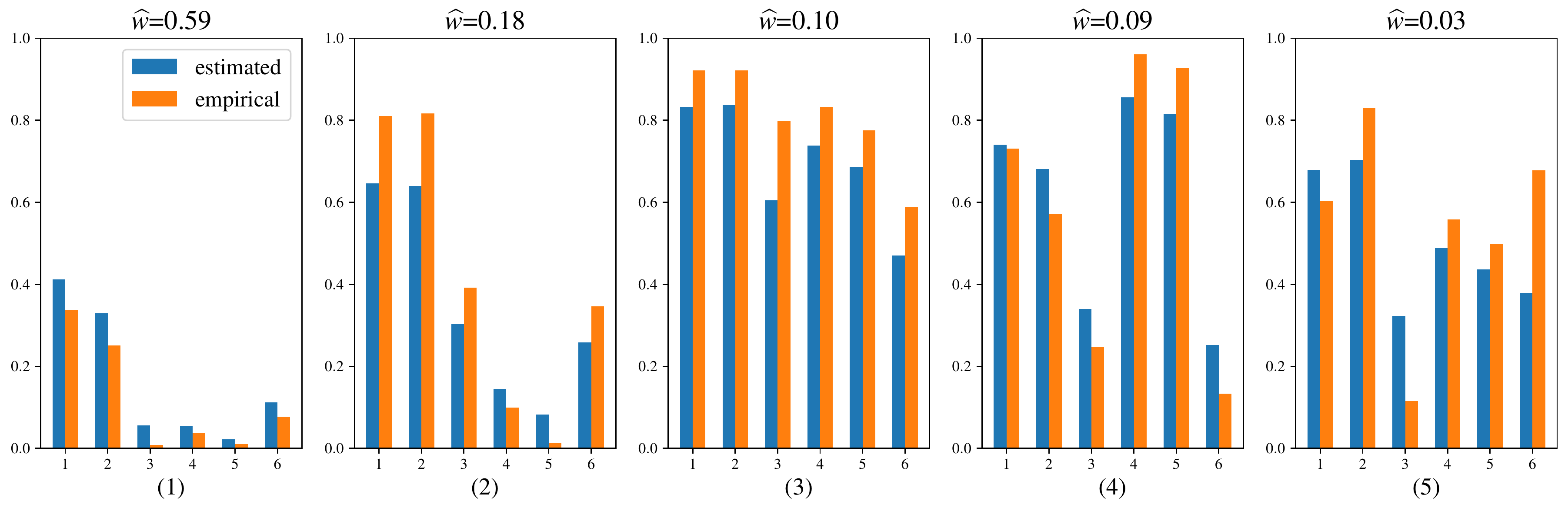}
    \caption{Estimated cluster centres $\hat \mu^{(a)}_h$ (in blue) and empirical estimates $\mu_h^{{\mathrm{emp}}}$ (in orange) when the clusters are
    sorted according to cluster sizes as given by the estimated weights $\hat w_h$, $h=1,\ldots,5$ (specified at the top of each plot). The clusters are also labelled  as $(1), \ldots, (5)$ (specified at the bottom of each plot).}
    \label{fig:latent_class_tau}
\end{figure}

We obtained
$P(k = 5 \mid \mbox{data}) \approx 1$.
As usually done in Bayesian mixture modelling, we
computed a point estimate of the latent partition of the data (as given by the unknown $c_i$'s)
by selecting, among the partitions visited during the MCMC iterations, the minimum point of the Binder loss function with equal misclassification cost, cf.\ \cite{binder1978bayesian}. 
Then, we evaluated the weights in each cluster by $\hat w_h = \# \hat C_h/ n$, $h=1,\ldots,5$,
where $\hat C_h$ is the estimated index set of data in cluster $h$. 
Furthermore, as in \cite{molitor2010bayesian}, we estimated the cluster centres 
by
\[
    \hat \mu^{(a)}_h = \frac{1}{M} \sum_{j=1}^M \frac{1}{\#\hat C_h } \sum_{i \in  \hat C_h} \mu^{j}_{c_i},\qquad h=1,\ldots,5.
\]
Figure~\ref{fig:latent_class_tau} shows these estimates, together
with the empirical frequencies
in each cluster as given by
\[
    \mu_h^{{\mathrm{emp}}} = \frac{1}{\#\hat C_h} \sum_{i \in  \hat C_h} y_i,\qquad h=1,\ldots,5.
\]
Note that in Figure~\ref{fig:latent_class_tau}, the estimated clusters are labeled $(1), \ldots, (5)$ and ordered by the estimated weights. 

The following interpretation of the clusters is consistent with the one given in \cite{li2018bayesian}: Figure~\ref{fig:latent_class_tau} shows that cluster (1) accounts for 59\% of the data and groups teenagers with few problematic behaviours, since all estimated and empirical cluster centers in the leftmost panel in Figure~\ref{fig:latent_class_tau} are small. Further,
cluster (2) groups 18\% of the subjects and describes minor problematic behaviours (relating to the first and second survey items).
Finally, clusters (3), (4), and (5) represent smaller groups of teenagers who are truly problematic, as
their tendency to commit small crimes (cluster 4) or fights (clusters 3 and 5) 
is very high. 

In Figure~\ref{fig:latent_class_tau}, there are discrepancies between the empirical frequencies and our estimates, 
see for instance the estimates of $\mu_{h1}$ and $\mu_{h2}$ in cluster (2) and of $\mu_{h3}$ and $\mu_{h6}$ in cluster (5).
These discrepancies can be explained by the use of the repulsive prior, which encourages separation among clusters.

\begin{figure}[t]
\centering
\includegraphics[width=0.5\linewidth]{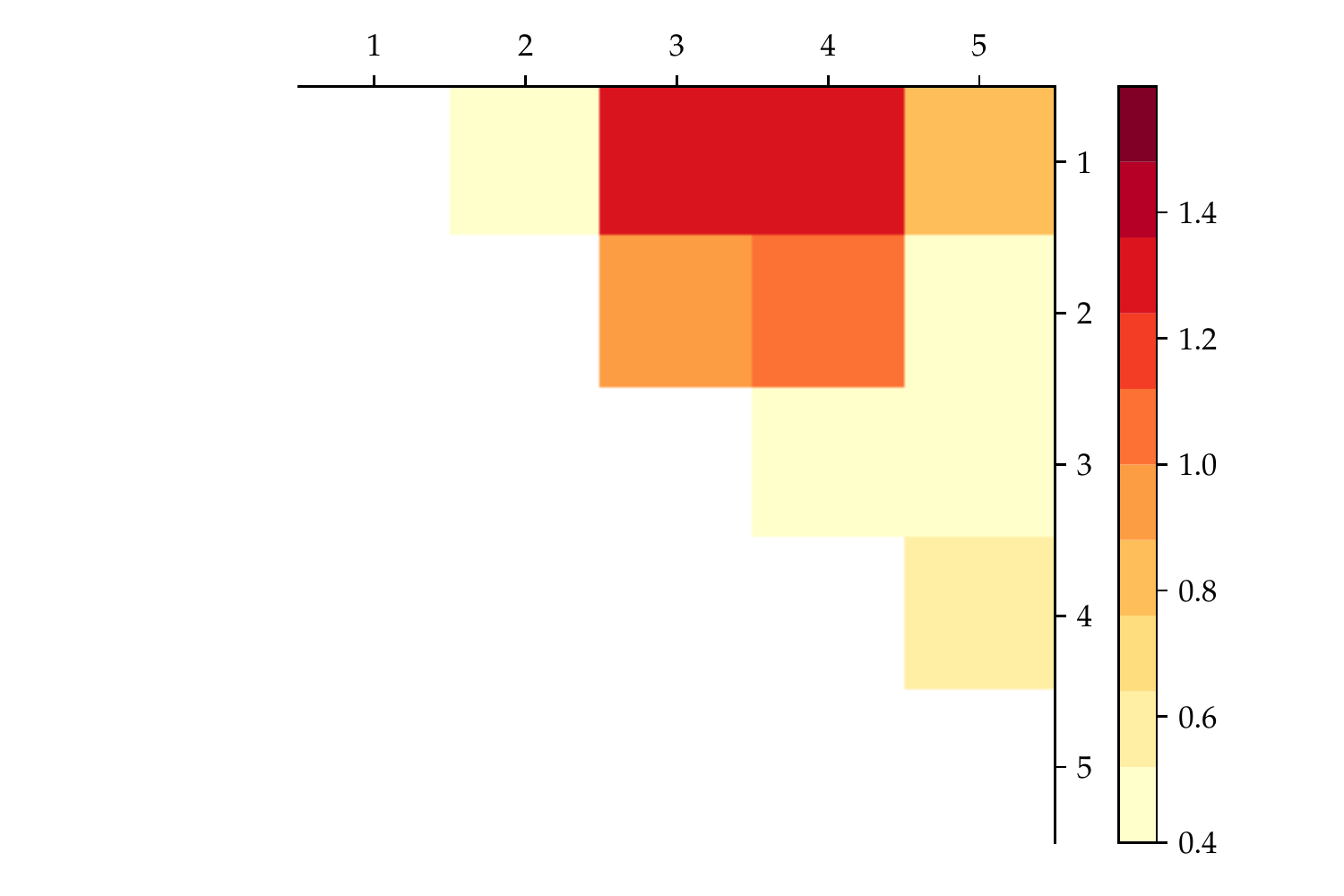}
\caption{Pairwise distances between the estimated $\hat \mu^{(a)}_h$, $h=1,\ldots,5$.}
\label{fig:tau_pdist}
\end{figure}

Moreover, Figure~\ref{fig:tau_pdist} shows the pairwise Euclidean distances among the estimated $\hat \mu^{(a)}_h$, $h=1,\ldots,5$.
Here, the smallest distance is around $0.41$, which is
close to the value of $\delta$  (which we fixed to be equal to $0.4$). Note that $\hat \mu^{(a)}_5$ is very
close to both $\hat \mu^{(a)}_2$ and $\hat \mu^{(a)}_3$; and $\hat \mu^{(a)}_1$
and $\hat \mu^{(a)}_2$ are close as well.

Finally, we performed posterior inference with
$\delta=0.5$ and $n^* = 100$ to induce more 
separation. 
In this case, our inference gave  four estimated clusters, in agreement with \cite{li2018bayesian}. 
However, compared to Figure~\ref{fig:latent_class_tau},
the  estimated cluster centres $\hat \mu^{(a)}_h$ were then further more different from the empirical frequencies $\mu_h^{{\mathrm{emp}}}$.  As noticed in Section~\ref{sec:simulation},
 this trade-off between density versus cluster estimation accuracy is  not surprising.

\section{Discussion}
\label{sec:discussion}

In this work we have contributed to the fast-growing literature on repulsive mixture models.
A main contribution is the  introduction of a unifying framework
which encompasses previously proposed repulsive mixtures as special cases.
In our setting, a repulsive point process is assumed as prior for `cluster centres' of the parametric kernel densities, thus making it more likely having a small number of well separated
clusters in the mixture model. In particular, we have showed the usefulness of the Strauss process prior, which is a simple example of a repulsive pairwise interaction point process.

By studying posterior characterization of the repulsive point process,
we were able to derive a Metropolis-within-Gibbs sampler that avoids
the arduous choice of problem-specific reversible jump proposals 
\citep{xu2016bayesian, bianchini2018determinantal}
and the computationally expensive evaluation of infinite summations
and integrals over the parameters space \citep{xie2019bayesian} 
 as seen in previous work.
When deriving the  posterior distribution of the repulsive point process prior, we extended the approach in \cite{argiento2019infinity} but framing our model within the class of normalized point processes mixture models.

Our MCMC algorithm can also handle cases
when the point process density involves an intractable normalizing constant, 
which has not been considered in the previous literature. 
In particular, we used
an ancillary variable method which eliminates the problem of having a ratio of normalizing
constants in the Hastings ratio when making posterior simulations for full conditional of the
hyperparameter. Since our mixture model is parsimonious (i.e., the number of components is typically small), the ancillary variable method relying on a perfect simulation algorithm is fast.

We tested our approach by extensive simulation studies, comparing it
to the reversible jump approach of \cite{xu2016bayesian} and \cite{bianchini2018determinantal}, where we concluded that our\change{ Metropolis-within-Gibbs (M-w-G) sampler}
has better mixing.
Our \change{M-w-G sampler} scales well with data dimension and this feature was particularly evident when we assumed the Strauss process as a prior for the cluster centers.
Furthermore, since repulsive mixture models encourage a small number of well separated components, thus controlling the computational cost, our 
algorithm was shown to scale well with sample size too. 

Finally, we illustrated  the advantages of repulsive mixtures against
the popular Dirichlet process mixtures and finite mixtures.
We concluded that repulsive mixtures are especially useful when the model is misspecified.

Several further extensions are possible. Beyond mixture models for cluster detection,  feature allocation problems and regression settings could be considered. 
 Further, adapting our approach to hierarchical and nested settings, where multiple groups of data are present, could be of interest.
Finally, extensions of our model to handle extremely high dimension data are also of interest, for instance in the field of genomics, where a repulsive prior would help in deriving interpretable results characterized by few and well separated clusters. 

\FloatBarrier	

\appendix

\section{Further details on the Metropolis-within-Gibbs sampler used for posterior simulation}\label{sec:extra_mcmc}

This section provides additional details for the Metropolis-within-Gibbs (M-w-G) sampler in Section~\ref{sec:algorithm}.

\subsection{The choice of the proposal distribution}\label{sec:proposal}

For most choices of the point process density $p(\bm\mu\mid\xi)$ and the mixture kernel $k(\cdot \mid \cdot)$, the update of the allocated means $\mu_h^{(a)}$ requires sampling from an unnormalized distribution, which we do via a Metropolis-Hastings step.
As proposal distribution we use a mixture of two normal distributions with means equal to the current value of $\mu_h^{(a)}$ but with different variances so that
\begin{equation}\label{e:proposal}
 p(\mu^\prime; \mu_h^{(a)}) = \kappa \mathcal{N}(\mu^\prime \mid \mu_h^{(a)}, \underline \sigma^2 I) + (1-\kappa) \mathcal{N}(\mu^\prime \mid \mu_h^{(a)}, \overline \sigma^2 I),
\end{equation}
where $\kappa = 0.9$, $\underline \sigma = 0.1$, and $\overline \sigma = 1.5$ when $q=1, 2$ and $\overline \sigma = 1.5q$ when $q > 2$.
The intuition that led us to consider such a proposal is as follows, where for ease of notation we drop the superscript $(a)$ when considering a current value of $\mu_h^{(a)}$, denoted $\mu_1$, and another cluster centre $\mu_2$. Suppose that
$\mu_1$ and $\mu_2$ are close and far from the remaining points in $\bm \mu$. 
If the number of observations allocated to $\mu_1$ is small, we want a proposal distribution $p(\mu_1^\prime; \mu_1)$ that
gives significant mass to values that are far from $\mu_2$, so that, given the repulsiveness of the point process, this proposal is likely to be accepted. This is the case when we sample from the second component of \eqref{e:proposal} (in fact, if $\mu_1^\prime$ is far from $\mu_1$, with sufficiently large probability it is far from $\mu_2$ as well).
On the other hand, if the number of observations allocated to $\mu_1$ is large, we want a proposal that gives significant mass to a neighborhood of of the current value of $\mu_1$, to get a precise fit of the data. This is what happens if we sample from the first component of \eqref{e:proposal}.

For the second component in \eqref{e:proposal}, instead of fixing $\overline{\sigma}$ as we do, \change{an alternative} is to exploit the properties of $g(\cdot\mid\xi)$ as follows. 
Suppose we condition on sampling from $\mathcal{N}(\mu_1^\prime \mid \mu_1, \overline \sigma^2 I)$ in \eqref{e:proposal}. Then $\|\mu_1^\prime - \mu_1 \|^2 / \overline \sigma^2 \sim \chi^2(q)$, the chi-squared distribution with $q$ degrees of freedom. 
Considering the Strauss density, a possibility is to fix $\overline{\sigma}$ to give sufficiently high mass to values of $\mu_1^\prime$ that are outside the range of interaction of $\mu_1$, i.e., such that $P(\|\mu_1^\prime - \mu_1 \|^2 > \delta) > p_0$ for some fixed $p_0$, with the intuition that this gives a positive probability to $\mu_1^\prime$ being distant at least $\delta$ also from $\mu_2$.
Considering the DPP density instead,
the same argument holds but replacing $\delta$ with the range of correlation $r_0$, cf.\ \cite{lavancier2015determinantal}. That is, \eqref{e:C} implies that $C$ is of the form $C(\mu_1, \mu_2)  = C_0(r)$ with $r=\|\mu_1-\mu_2\|$, and defining the corresponding correlation function $R(r) = C_0(r) / C_0(0)$, $r_0$ is chosen such that $R(r)$ is effectively zero.

\subsection{The exchange algorithm and perfect simulation}\label{sec:exchange}

With the same notation as Section~\ref{sec:xi}, the exchange algorithm \citep{murray2006mcmc} consists of the following steps:
\begin{enumerate}
\item Propose $\xi^\prime \sim p(\xi^\prime; \xi)$.
\item Generate an auxiliary variable $\bm \mu^{\text{aux}} \sim g(\bm \mu \mid \xi^\prime) / Z_{\xi^\prime} \propto g(\bm \mu \mid \xi^\prime)$.
\item Accept $\xi^\prime$ with probability $\min\{1, \alpha^*\}$ where
\[
    \alpha^* \equiv \alpha^*(\xi; \xi^\prime \mid \cdots) = \frac{p(\xi^\prime) g(\bm \mu \mid \xi^\prime) p(\xi; \xi^\prime)}{p(\xi) g(\bm \mu \mid \xi) p(\xi^\prime; \xi)} \times \frac{g(\bm \mu^{\text{aux}} \mid \xi)}{g(\bm \mu^{\text{aux}} \mid \xi^\prime)}.
\]
\end{enumerate}
Comparing $\alpha^*$ to the acceptance ratio in \eqref{e:acc-jm}, note that the ratio $Z_\xi / Z_{\xi^\prime}$ has been replaced by a ratio of unnormalized densities, evaluated in the auxiliary variable $\bm \mu^{\text{aux}}$.
The main difficulty is sampling $\bm \mu^{\text{aux}}$, which must follow the distribution of $\bm \mu$ given $\xi^\prime$.
To this end, we employ the stochastic dominated coupling from the past algorithm in \cite{kendall2000perfect}, which extends the coupling from the past algorithm in \cite{propp1996exact} to uncountable partially ordered spaces.
Specifically, we employed in our code Algorithm 11.7 in \cite{moller2003statistical}.

\section{Additional simulation studies}\label{sec:extra_sim}

In addition to the simulation studies in Section~\ref{sec:simulation}, below we discuss different aspects of the M-w-G sampler and posterior inference.

\subsection{Comparison of run-times and posterior inference when using DPP and Strauss process priors}\label{sec:strauss_vs_dpp}

For $q=1, 2, \ldots, 5$, we  simulated $n=200$ observations from \eqref{eq:dgp2} with $\mu_0 = (-5, \ldots, -5)$, $\Sigma_0 = I_q$, $\omega=1$, $\mu_1 = 5$, and $\sigma_1=1$. Then we
 applied our M-w-G sampler when the marginal prior for $\bm \mu$ is either the DPP or the Strauss process, with hyperparameters as in Section~\ref{sec:prior}. 
Here, we considered two truncation levels for the approximation of the 
DPP density in \eqref{e:C}, namely $N=5$ and $N=10$ (for comparison, \cite{bianchini2018determinantal} suggested $N=50$ when $q=1$).

\begin{figure}[t]
\centering
\includegraphics[width=0.5\linewidth]{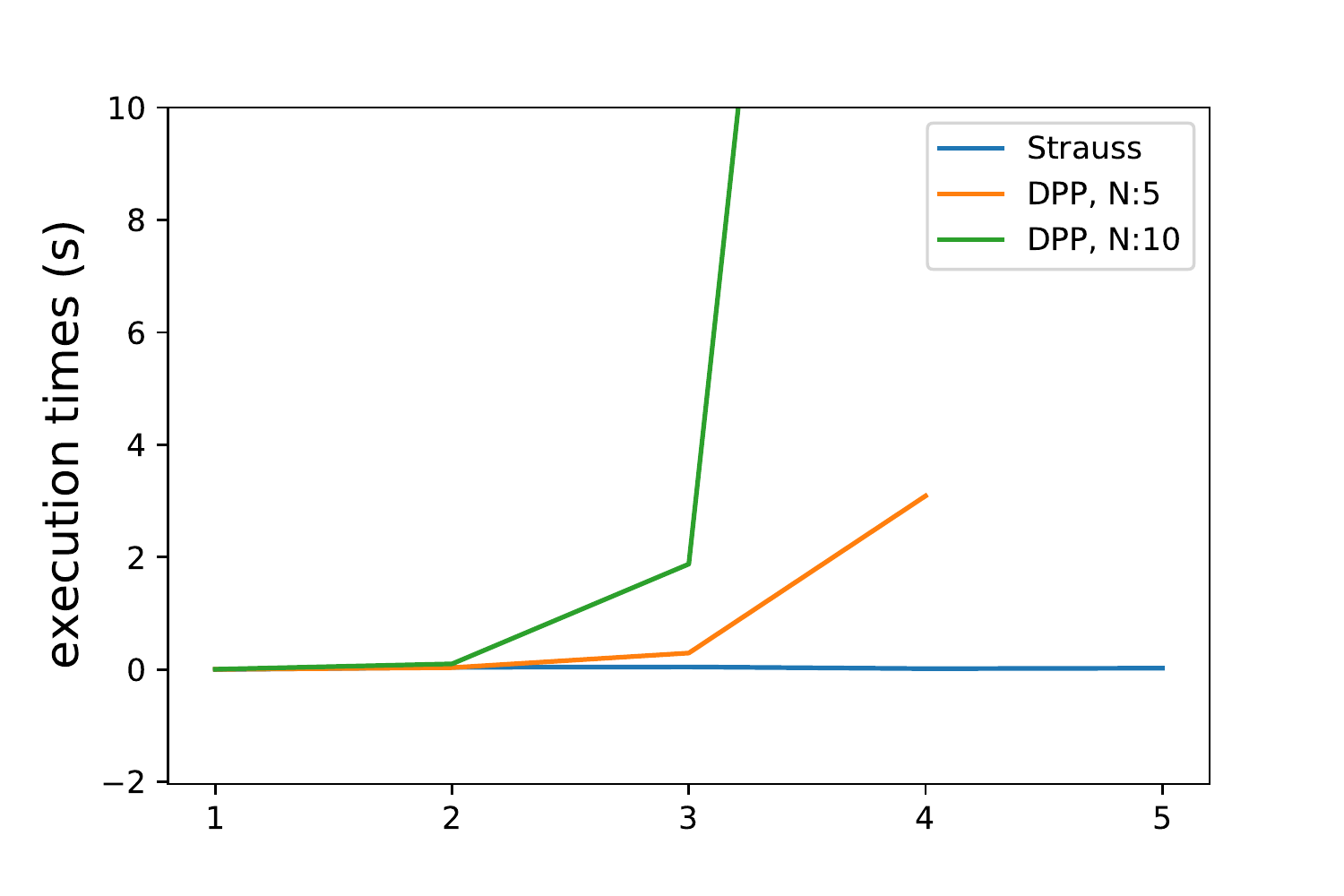}
\caption{Per-iteration run-times as a function of data dimension $q$ in case of DPP (with truncation levels $N=5$ or $10$) and Strauss process priors for $\bm \mu$.
}
\label{fig:runtimes}
\end{figure}

Figure~\ref{fig:runtimes} shows the per-iteration run-times of the M-w-G sampler as a function of the dimension $q$ under either the DPP or Strauss process prior for $\bm\mu$.
 For each value of
$N$, the computational cost associated to the DPP grows exponentially fast as the dimension $q$ increases,
unlike in the case of the Strauss process. In fact,
the unnormalized density of the Strauss process is almost immediate to compute, and 
since the Strauss prior is quite informative on the number of components, cf.\ Section~\ref{sec:strauss_params}, the perfect simulation algorithm (see Section~\ref{sec:xi}) does not impact significantly on the computational cost.
Although not appreciable from Figure~\ref{fig:runtimes}, the computational cost of our algorithm increases
significantly with data dimension $q$ also when we consider the Strauss process;
in this case, the per-iteration computational cost goes from 0.0016 sec when $q=1$ to 0.07 sec when  $q=5$, i.e., it increases by a factor of roughly 50.

\begin{figure}[t]
\begin{subfigure}{0.5\linewidth}
    \centering
    \includegraphics[width=\linewidth]{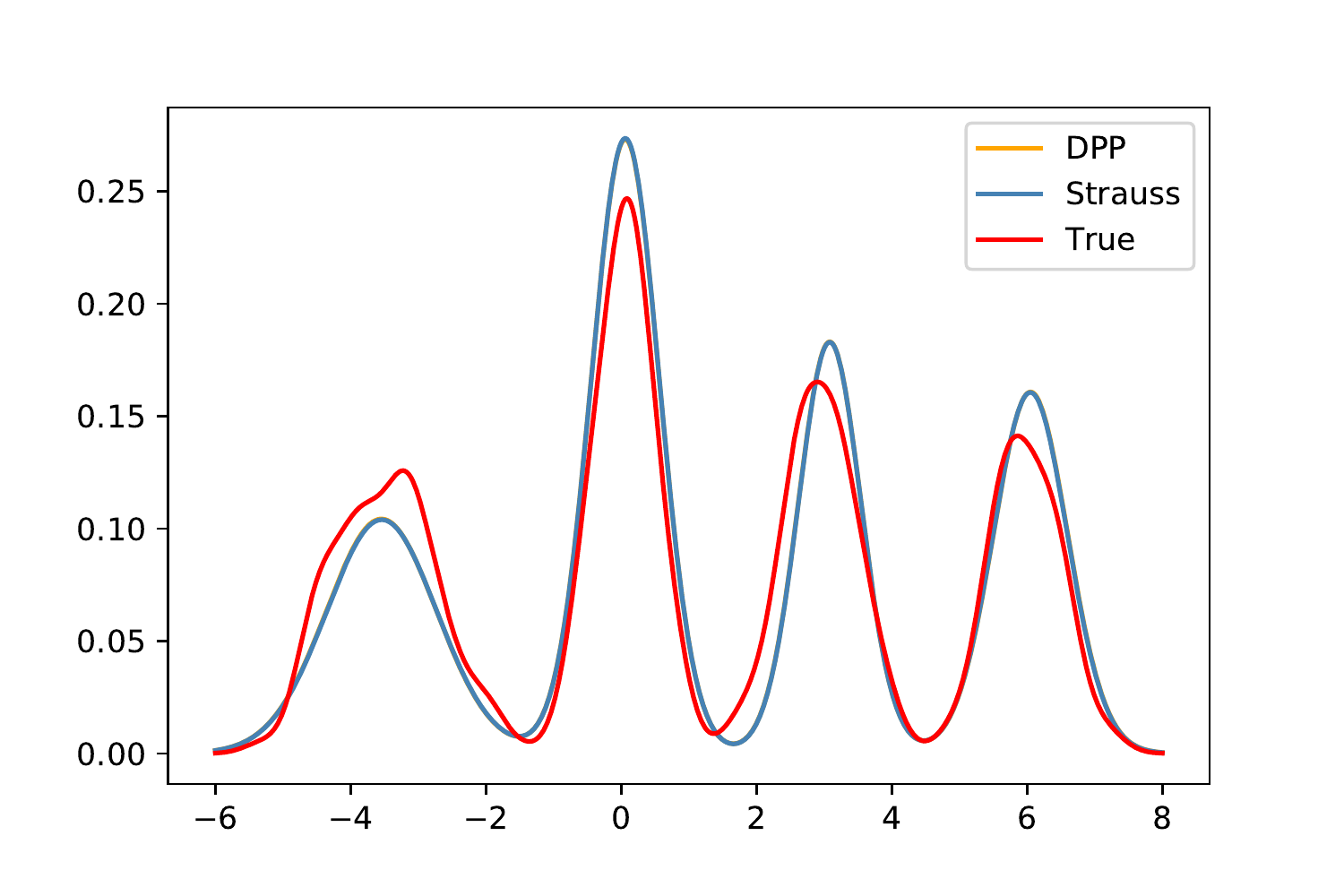}
\end{subfigure} %
\begin{subfigure}{0.5\linewidth}
    \centering
    \includegraphics[width=\linewidth]{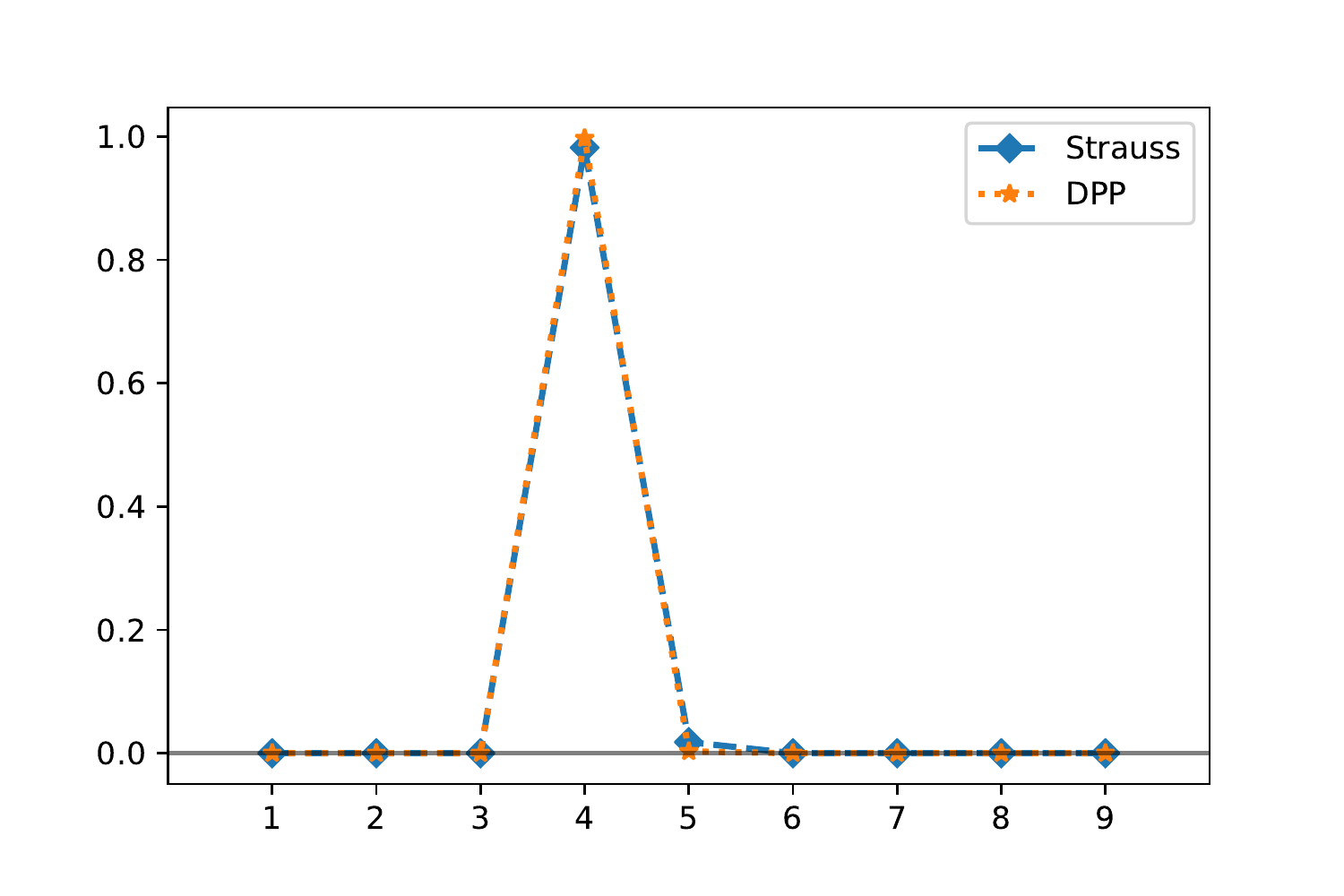}
\end{subfigure}
\caption{Bayesian mixture density estimates (left) and posterior distributions of the number of clusters (right) under
the Strauss process (blue lines) and DPP (orange lines) priors for $\bm\mu$, together with the true mixture density which has four components.
The orange lines overlap almost perfectly with the blue lines so that they are hardly visible.}
\label{fig:dpp_vs_strauss}
\end{figure}

As a further comparison, we simulated $500$ univariate observations from model~\eqref{eq:dgp1} and made again posterior computations under  the Strauss process or the DPP prior for $\bm\mu$, where for the DPP density we fixed $\beta = 10$ (corresponding to the highest ESS in Table~\ref{tab:vs_rj}). 
For both cases of prior models, we ran the M-w-G sampler for $100,000$ iterations discarding the first $50,000$ as a burn-in and keeping one every ten iterations, for a final sample size of $5,000$.
Figure~\ref{fig:dpp_vs_strauss} shows the true data generating density, together with  Bayesian mixture density estimates and posterior distributions of the number of clusters under the two point process priors.
Note that the two density estimates, as well as the two posterior distributions of the number of clusters, overlap almost perfectly.
The Strauss process seems a good choice to model the prior of $\bm\mu$ since it, for  a much smaller computational cost, provides  same posterior summaries as the DPP.

\subsection{Accuracy of cluster estimates}
\label{sec:clust}

\begin{figure}
  \centering
  \includegraphics[width=0.9\linewidth]{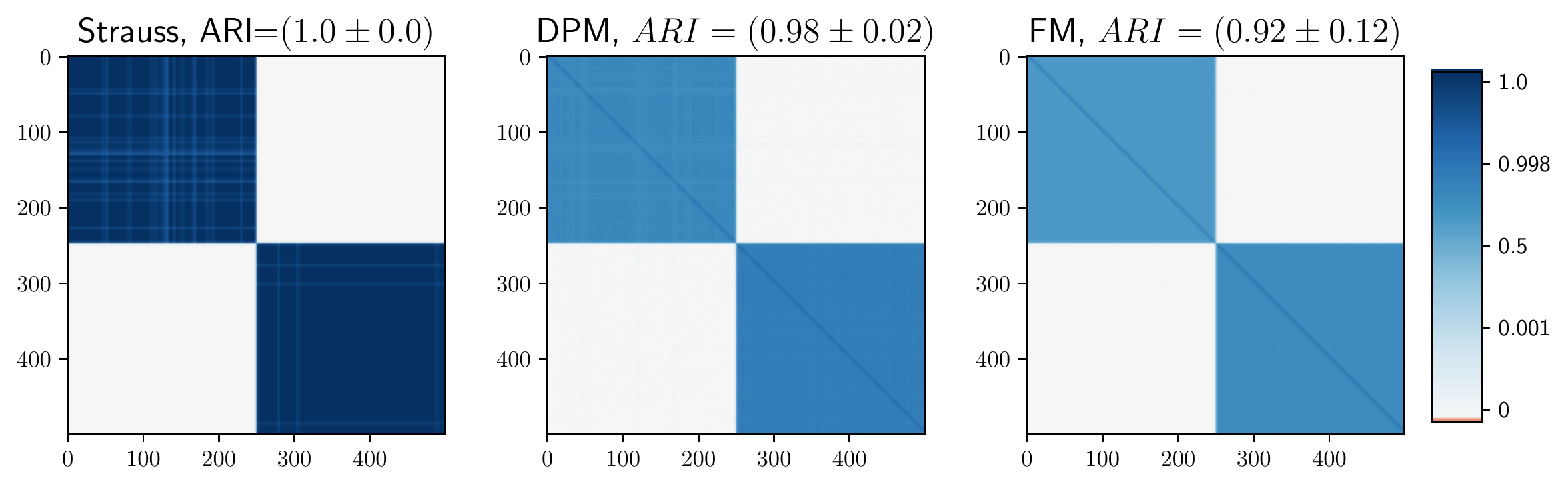}
  \caption{Posterior similarity matrices  and ARI scores under the three models for the mixture of the univariate $t$ and skew-normal distributions discussed in Section~\ref{sec:vs_dpm_mfmf}.
  The colors are on a logit scale to highlight differences around one.}
  \label{fig:psm}
\end{figure}

Figure~\ref{fig:psm} shows the posterior similarity matrices and the Adjusted Rand Index (ARI) scores for the univariate mixture of $t$ and skew-normal distribution discussed in Section~\ref{sec:vs_dpm_mfmf}.
The ARI is computed from the cluster labels $\bm c$ at each iteration of the MCMC chain as a measure of similarity between the estimated clusters and the true cluster. It is bounded by 1 and the larger value it assumes, the more similar is the estimated cluster to the true one. We report the posterior mean  of the ARI $\pm$ one standard deviation on top of each posterior similarity matrix in Figure~\ref{fig:psm}.
The difference in the posterior similarity matrices is not so pronounced, but our repulsive mixture model gives the best ARI.

\subsection{The effect of the number of clusters}

\begin{figure}[t]
\begin{subfigure}{0.5\linewidth}
    \centering
    \includegraphics[width=\linewidth]{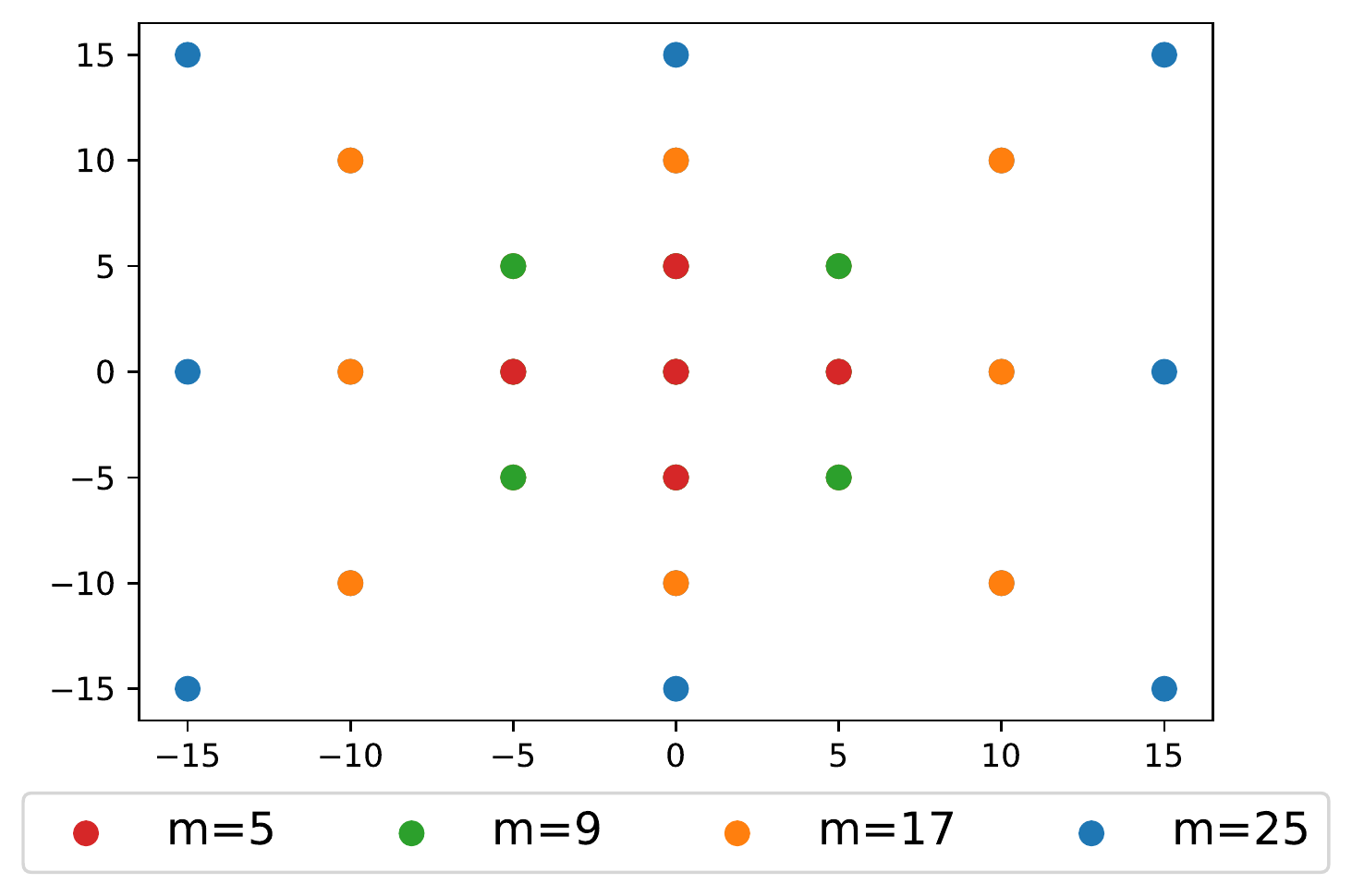}
\end{subfigure} %
\begin{subfigure}{0.5\linewidth}
    \centering
    \includegraphics[width=0.95\linewidth]{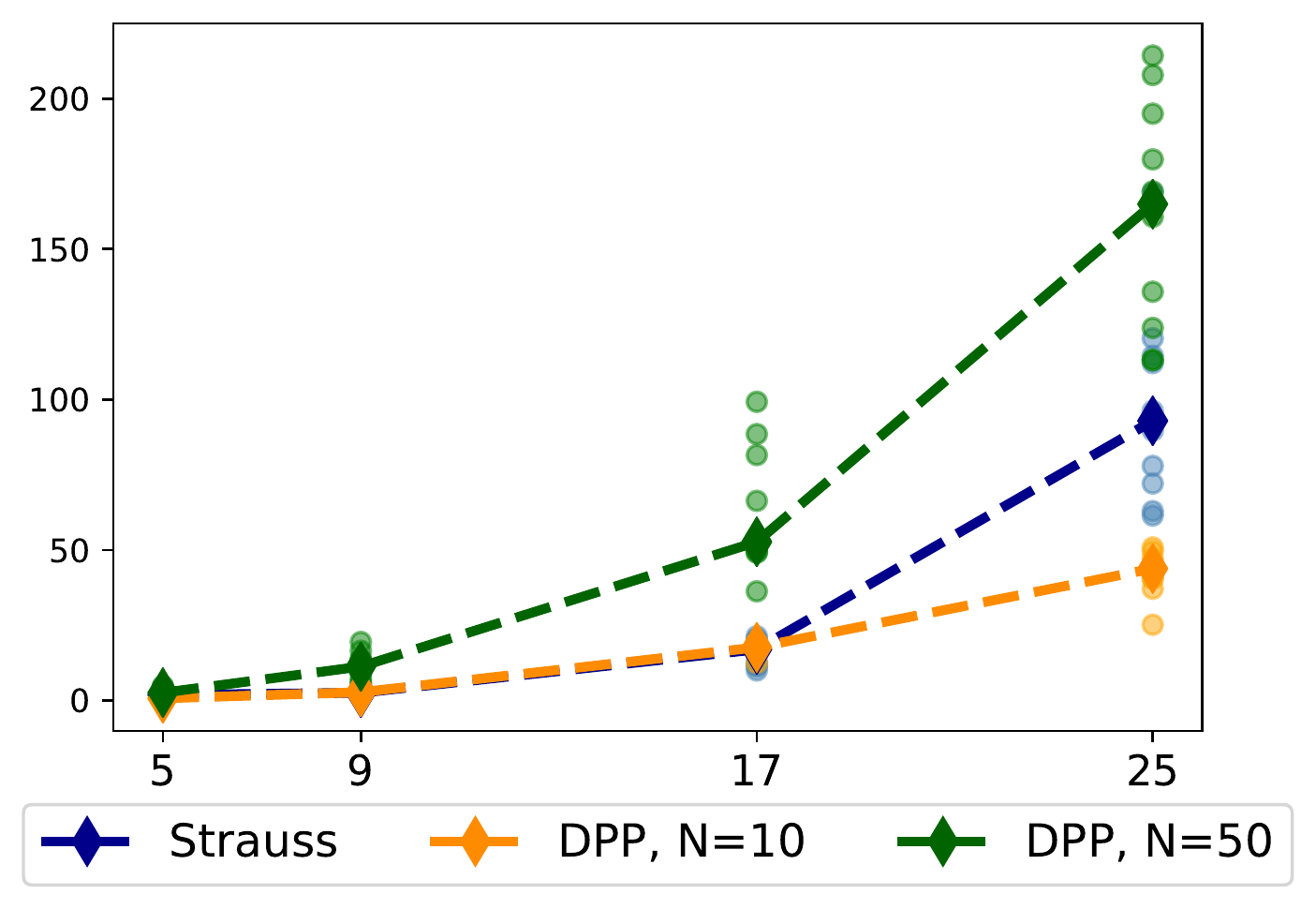}
\end{subfigure}
\caption{Locations of the true data generating process (left) and run-time comparison (right). The plot of the locations should be intended as follows: for $m=5$ only the points labelled accordingly are considered, for $m=9$ the points labeled as $m=5$ and $m=9$ are considered and so on. 
The run-times (in seconds) over 50 independently simulated datasets for each value of $m$ are denoted by dots, we also report the median times as diamonds with a dashed line connecting them.
} 
\label{fig:inc_m}
\end{figure}

We consider how the number of clusters affects the performance of our M-w-G sampler.
When $\bm \mu$ is distributed as the Strauss process, at every step of the MCMC algorithm a perfect simulation of $\bm \mu$ is required.
The perfect simulation algorithm we use has a finite but random computational cost and, as argued in Section~\ref{sec:xi}, it might become infeasible for a large number of clusters.
On the other hand, when $\bm \mu$ is a DPP, the approximation of its density requires computing the determinant of the matrix $C^\prime$ in \eqref{e:unnormdppdensity}, which scales cubically with $m$. Furthermore, for the specific DPP considered in \eqref{e:C} computing $C^\prime$ requires the evaluation of
$O(N^q m^2)$ inner products.

We generated $n=500$ observations from a mixture of $m=5, 9, 17, 25$ bivariate Gaussian densities, with locations given in Figure~\ref{fig:inc_m} (left), equal covariance matrices given by $0.5 I_2$, and with equal mixture weights.
We compared the run-times (in seconds) required to complete 200 iterations with our M-w-G sampler when $\bm \mu$ is distributed either as the Strauss or the determinantal point process. Prior hyperparameters are fixed as in Section~\ref{sec:prior} (with $M_{\max} = 5 m$) and Section~\ref{sec:simulation}. For the DPP, we considered two truncation levels of the spectral density, $N=10, 50$.
For each choice of $m$ we generated 50 independent datasets and for the 200 M-w-G sampler iterations we used fixed and different independent random seeds.

In Figure~\ref{fig:inc_m} (right) for each $m$ the run-times over the 50 independent datasets are denoted by dots, the median times by diamonds, and the median times are connected by a dashed line.
We see that the DPP with $N=50$ is the most computationally demanding model for all values of $m$. 
When $m=5, 9$, the Strauss process is significantly faster (up to 10 times faster) than the DPP with $N=10$; instead, when $m=17$, they have comparable computational costs.
When $m=25$, the perfect simulation algorithm starts to become more demanding; for example, the computational cost for the Strauss process is almost twice the one for the DPP with $N=10$.

\subsection{The effect of the data dimension}\label{sec:large_m}

Below we compare our repulsive mixture model, the finite mixture model (FM) in \cite{argiento2019infinity}, and the Dirichlet process mixture model (DPM). See Section~\ref{sec:vs_dpm_mfmf} for further details on how posterior inference is performed under the different models.
In particular, we fix the hyper-parameters according to Sections~\ref{sec:prior} and \ref{sec:simulation}.

If the sample size $n$ is not significantly larger than the data dimension $q$, the use of commonly employed MCMC algorithms may be problematic for the following reasons. If  $k(\cdot \mid \cdot)$ is a multivariate Gaussian density with non-zero correlations, the number of parameters to be estimated is much larger than $n$.
Further, the curse of dimensionality, common to all clustering problems \citep{kriegel2009clustering}, implies a poor mixing of the algorithms.
\change{In addition to that}, when considering a repulsive mixture model, things might be further complicated by either
the need of perfect simulation to update possible hyperparameters $\xi$ (when $\bm \mu$ follows the Strauss process) or the computation of the spectral density (when $\bm \mu$ follows the DPP given by \eqref{e:C}) which becomes prohibitive even for moderate values of $q$, as shown in Figure~\ref{fig:runtimes}.
Therefore, below we consider only the Strauss process and perform a simulation to assess the performance of repulsive  versus non-repulsive mixtures when
 $q=2, 5, 10, 15, 20, 25, 30$ increases. Moreover, we simulated $n=200$ observations from
\[
    y_i \iid 0.5 \calN( -5/\sqrt{q} \bm 1_q, I_q) + 0.5 \calN( 5/\sqrt{q} \bm 1_q, I_q)
\]
where $\bm 1_q$ denotes the vector in $\mathbb{R}^q$ with elements all equal to one. 
\begin{table}
\centering
\begin{tabular}{c | c | c c c c c c}
\multicolumn{2}{c|}{} & $q=5$ & $q=10$ & $q=15$ & $q=20$ & $q=25$ & $q=30$ \\
\hline
\multirow{3}{*}{Strauss} & ARI & 1.0 & 1.0 & 1.0 & 1.0 & 1.0 & 1.0 \\
& ESS & 240.3 & 250.1 & 0.0 & 0.0 & 0.0 & 0.0  \\
& $\mathbb{E}[k \mid \text{data}]$ & 2.01 & 2.005 & 2.0 & 2.0 & 2.0 & 2.0 \\
\hline
\multirow{3}{*}{FM} & ARI & 1.0 & 1.0 & 1.0 & 1.0 & 1.0 & 1.0 \\
& ESS & 7.4 & 0.0 & 0.0 & 0.0 & 0.0 & 0.0  \\
& $\mathbb{E}[k \mid \text{data}]$ & 2.01 & 2.0 & 2.0 & 2.0 & 2.0 & 2.0 \\
\hline
\multirow{3}{*}{DPM} & ARI & 1.0 & 0.0 & 0.0 & 0.0 & 0.0 & 0.0 \\
& ESS & 0.0 & 0.0 & 0.0 & 0.0 & 0.0 & 0.0 \\
& $\mathbb{E}[k \mid \text{data}]$ & 2.00 & 1.0 & 1.0 & 1.0 & 1.0 & 1.0 
\end{tabular}
\caption{Adjusted Rand Index (ARI), effective sample size for the chain of the number of clusters $k$ and posterior mean of $k$ under the repulsive mixture model (Strauss), the non repulsive finite mixture model (FM) and the Dirichlet process mixture model (DPM).}
\label{tab:inc_q}
\end{table}

Table~\ref{tab:inc_q} reports posterior summaries as $q$ increases for
the three models.
MCMC chains were run for $11,000$ iterations discarding the first $10,000$ as burn-in, so that the effective sample size must be referred to a total number of MCMC iterations equal to $1,000$. It is clear from Table~\ref{tab:inc_q} that as $q$ increases, the mixing of the chains becomes progressively worse for all the models.
In particular, the table shows the effective sample size (ESS) for the three cases: 
For our repulsive mixture model, the number of clusters $k$ is constant for all the MCMC iterations when $q \geq 15$, and so ESS is zero; for FM, the ESS is zero when $q \geq 10$;  and for DPM the ESS is zero for all values of $q$.
The difference in the ARI scores is simply explained by the different strategy of initialization of the different software we ran: In our code for the M-w-G sampler, observations are initially randomly subdivided into 10 clusters; in the package \texttt{AntMAN}, which we used to fit the FM model, one cluster per observation is created; in the package \texttt{BNPMix}, used to fit the DPM model, all observations are initially allocated to one single cluster. 
In the latter case, the proposal of a new cluster is never accepted.
Using our software or the package \texttt{AntMAN} instead, after a few MCMC iterations the observations are (correctly) partitioned into $k=2$ clusters and no additional cluster is ever created.

Considering the effective sample size of $k$, Table~\ref{tab:inc_q} shows that repulsive mixture models might offer an advantage over non-repulsive mixture models when $q \leq 10$.
We believe that the poor performance of FM and DPM is due to prior assumptions for the following reason. Note that both models assume that the parameters $\{(\mu_h, \gamma_h)\}_h$ are a priori iid and normal inverse-Wishart distributed, with $\mathbb{E}[\mu_h] = 0$. Thus, 
as $q$ increases, the multivariate Gaussian distribution becomes more and more concentrated around the mean, due to the so-called curse of dimensionality, so that proposing a new value for $\mu_h$ from the prior that is near to any of the observations becomes less likely.
Instead, when considering a Strauss point process as prior for $\bm \mu$, the proposed means are not concentrated around the origin, which led to a better mixing when $q=5, 10$. When $q \geq 15$, we believe that the volume of the rectangle containing all observations becomes so large that also the repulsive mixture models suffer from the curse of dimensionality. 

Perfect simulation is not a bottleneck here, as the number of points in the Strauss process is small. 
However, in one of several independent simulations, an unlucky initialization led to  a large value of $m$ in the first few iterations. 
As a consequence, the perfect simulation algorithm took longer to coalesce and indeed caused an out-of-memory problem on a 32 GB laptop. 

Finally, when $q \rightarrow +\infty$, \cite{chandra2020escaping} show how
the posterior distribution under non repulsive mixture models either assigns all the observations to the same cluster or each observation to a separate cluster.
These authors propose to consider mixtures in a latent space to overcome such issue, similarly to \cite{ghahramani1996algorithm}.
Extensions of latent mixture models to account for repulsiveness are currently being investigated;  see \cite{ghilotti2021thesis}.

\section{Removing the rectangular support assumption}\label{sec:support}

Often we have assumed that the points of $\bm \mu$ have support given by a rectangular set $R$: For the theory in Sections~\ref{sec:prior_mu}--\ref{sec:algorithm}, we made that assumption only for specificity and simplicity;
in Section~\ref{sec:simulation}, we considered Gaussian mixture models and determined the rectangle $R$  from the observations; while in 
Section~\ref{sec:teenage}, we considered the multivariate Bernoulli kernel and $R = [0, 1]^q$. Apart from the case of a DPP prior, it is often easy to modify everything without assuming $R$ is rectangular and even compactness of $R$ may be not be needed, In fact, 
the birth-death Metropolis-Hastings algorithm, which we always use to simulate the non-allocated process $\bm \mu^{(na)}$, can be specified in a very general setting, see \cite{geyer1994simulation}. 
On the other hand, for a DPP prior, compactness of $R$ is needed when specifying a DPP density with respect to $\mathrm{d}\bm\mu$, and $R$ needs to be a rectangle in order to use the spectral approach discussed in \cite{lavancier2015determinantal}.
Recently, \cite{poinas2021asymptotic} proposed a novel approximation of a general DPP density that does not require $R$ to be rectangular (but still requires $R$ is bounded). 
	
\bibliographystyle{ba}
\bibliography{bib_dpp_new}

\end{document}